\PassOptionsToPackage{unicode}{hyperref}
\PassOptionsToPackage{hyphens}{url}
\documentclass[
]{scrartcl}
\usepackage{amsmath,amssymb}
\usepackage{iftex}
\ifPDFTeX
  \usepackage[T1]{fontenc}
  \usepackage[utf8]{inputenc}
  \usepackage{textcomp} 
\else 
  \usepackage{unicode-math} 
  \defaultfontfeatures{Scale=MatchLowercase}
  \defaultfontfeatures[\rmfamily]{Ligatures=TeX,Scale=1}
\fi
\usepackage{lmodern}
\ifPDFTeX\else
\fi
\IfFileExists{upquote.sty}{\usepackage{upquote}}{}
\IfFileExists{microtype.sty}{
  \usepackage[]{microtype}
  \UseMicrotypeSet[protrusion]{basicmath} 
}{}
\makeatletter
\@ifundefined{KOMAClassName}{
  \IfFileExists{parskip.sty}{%
    \usepackage{parskip}
  }{
    \setlength{\parindent}{0pt}
    \setlength{\parskip}{6pt plus 2pt minus 1pt}}
}{
  \KOMAoptions{parskip=half}}
\makeatother
\usepackage{xcolor}
\usepackage{color}
\usepackage{fancyvrb}

\DefineVerbatimEnvironment{Highlighting}{Verbatim}{commandchars=\\\{\}}
\usepackage{framed}
\definecolor{shadecolor}{RGB}{255,255,255}
\newenvironment{Shaded}{\begin{snugshade}}{\end{snugshade}}

\newcommand{\AttributeTok}[1]{\textcolor[rgb]{0.00,0.00,0.00}{#1}}

\newcommand{\DecValTok}[1]{\textcolor[rgb]{0.00,0.00,0.00}{#1}}

\newcommand{\FunctionTok}[1]{\textcolor[rgb]{0.00,0.00,0.00}{#1}}

\newcommand{\NormalTok}[1]{\textcolor[rgb]{0.00,0.00,0.00}{#1}}

\newcommand{\OtherTok}[1]{\textcolor[rgb]{0.00,0.00,0.00}{#1}}

\newcommand{\SpecialCharTok}[1]{\textcolor[rgb]{0.00,0.00,0.00}{#1}}

\newcommand{\StringTok}[1]{\textcolor[rgb]{0.00,0.00,0.00}{#1}}

\usepackage{longtable,booktabs,array}
\usepackage{calc} 
\usepackage{etoolbox}
\makeatletter
\patchcmd\longtable{\par}{\if@noskipsec\mbox{}\fi\par}{}{}
\makeatother
\IfFileExists{footnotehyper.sty}{\usepackage{footnotehyper}}{\usepackage{footnote}}
\makesavenoteenv{longtable}
\usepackage{graphicx}
\makeatletter
\newsavebox\pandoc@box
\newcommand*\pandocbounded[1]{
  \sbox\pandoc@box{#1}%
  \Gscale@div\@tempa{\textheight}{\dimexpr\ht\pandoc@box+\dp\pandoc@box\relax}%
  \Gscale@div\@tempb{\linewidth}{\wd\pandoc@box}%
  \ifdim\@tempb\p@<\@tempa\p@\let\@tempa\@tempb\fi
  \ifdim\@tempa\p@<\p@\scalebox{\@tempa}{\usebox\pandoc@box}%
  \else\usebox{\pandoc@box}%
  \fi%
}
\def\fps@figure{htbp}
\makeatother
\providecommand{\tightlist}{%
  \setlength{\itemsep}{0pt}\setlength{\parskip}{0pt}}
\makeatletter
\ifx\paragraph\undefined\else
  \let\oldparagraph\paragraph
  \renewcommand{\paragraph}{
    \@ifstar
      \xxxParagraphStar
      \xxxParagraphNoStar
  }
  \newcommand{\xxxParagraphStar}[1]{\oldparagraph*{#1}\mbox{}}
  \newcommand{\xxxParagraphNoStar}[1]{\oldparagraph{#1}\mbox{}}
\fi
\ifx\subparagraph\undefined\else
  \let\oldsubparagraph\subparagraph
  \renewcommand{\subparagraph}{
    \@ifstar
      \xxxSubParagraphStar
      \xxxSubParagraphNoStar
  }
  \newcommand{\xxxSubParagraphStar}[1]{\oldsubparagraph*{#1}\mbox{}}
  \newcommand{\xxxSubParagraphNoStar}[1]{\oldsubparagraph{#1}\mbox{}}
\fi
\makeatother
\makeatletter
\@ifpackageloaded{caption}{}{\usepackage{caption}}
\AtBeginDocument{%
\ifdefined\contentsname
  \renewcommand*\contentsname{Table of contents}
\else
  \newcommand\contentsname{Table of contents}
\fi
\ifdefined\listfigurename
  \renewcommand*\listfigurename{List of Figures}
\else
  \newcommand\listfigurename{List of Figures}
\fi
\ifdefined\listtablename
  \renewcommand*\listtablename{List of Tables}
\else
  \newcommand\listtablename{List of Tables}
\fi
\ifdefined\figurename
  \renewcommand*\figurename{Figure}
\else
  \newcommand\figurename{Figure}
\fi
\ifdefined\tablename
  \renewcommand*\tablename{Table}
\else
  \newcommand\tablename{Table}
\fi
}
\@ifpackageloaded{float}{}{\usepackage{float}}
\floatstyle{ruled}
\@ifundefined{c@chapter}{\newfloat{codelisting}{h}{lop}}{\newfloat{codelisting}{h}{lop}[chapter]}
\floatname{codelisting}{Listing}

\makeatother
\makeatletter
\@ifpackageloaded{caption}{}{\usepackage{caption}}
\@ifpackageloaded{subcaption}{}{\usepackage{subcaption}}
\makeatother
\usepackage[authoryear,round]{natbib}
\usepackage{bookmark}
\IfFileExists{xurl.sty}{\usepackage{xurl}}{} 
\hypersetup{
  pdftitle={Prior-Posterior Derived-Predictive Consistency Checks for Post-Estimation Calculated Quantities of Interest (QOI-Check)},
  pdfauthor={Holger Sennhenn-Reulen },
  pdfkeywords={Simulation Based Calibration (SBC), Posterior Predictive
Check, Reference Grid},
  hidelinks,
  pdfcreator={LaTeX via pandoc}}

\title{Prior-Posterior Derived-Predictive Consistency Checks for
Post-Estimation Calculated Quantities of Interest (\emph{QOI-Check})}
\usepackage{etoolbox}
\makeatletter
\providecommand{\subtitle}[1]{
  \apptocmd{\@title}{\par {\large #1 \par}}{}{}
}
\makeatother
\subtitle{A Preprint}

\usepackage{authblk}
\usepackage{orcidlink}

\author{Holger
Sennhenn-Reulen\textsuperscript{\orcidlink{0000-0002-4782-4387}}
\linebreak \textsuperscript{Northwest German Forest Research Institute (NW-FVA), Germany}}

\date{March 20, 2025}

\usepackage{lmodern}

\usepackage{tikz}
\usetikzlibrary{shapes.geometric, arrows}
\usepackage{subcaption}
\usepackage{enumitem}

\usepackage[draft, defaultcolor=teal]{changes}

\begin{document}
\setlist[description]{leftmargin = *} 
\maketitle
\begin{abstract}
\textsf{\textbf{Abstract}} With flexible modeling software -- such as
the probabilistic programming language \emph{Stan}
\citep{CarpenterEtAl2017} -- growing in popularity, \emph{quantities of
interest} (\emph{QOI}s) calculated post-estimation are increasingly
desired and customly implemented, both by statistical software
developers and applied scientists. Examples of QOI include the marginal
expectation of a multilevel model with a non-linear link function, or an
ANOVA decomposition of a bivariate regression spline. For this, the
\emph{QOI-Check} is introduced, a systematic approach to ensure proper
calibration and correct interpretation of QOIs. It contributes to
\emph{Bayesian Workflow} \citep{GelmanEtAl2020}, and aims to improve the
interpretability and trust in post-estimation conclusions based on QOIs.
The QOI-Check builds upon \emph{Simulation Based Calibration}
(\emph{SBC}) \citep{ModrakEtAl2023}, and the \emph{Holdout Predictive
Check} (\emph{HPC}) \citep{MoranBleiRanganath2023}. SBC verifies
computational reliability of Bayesian inference algorithms by
consistency check of posterior with prior when the posterior is
estimated on prior-predicted data, while HPC ensures robust inference by
assessing consistency of model predictions with holdout data. SBC and
HPC are combined in QOI-Checking for validating post-estimation QOI
calculation and interpretation in the context of a (hypothetical)
population definition underlying the QOI.
\end{abstract}

\clearpage

\section*{Notation}\label{notation}
\addcontentsline{toc}{section}{Notation}

\begin{itemize}
\tightlist
\item
  Observation unit index \(i=1,\ldots,N\)
\item
  Posterior sample index \(s=1,\ldots,S\)
\item
  Simulation run index \(r=1,\ldots,R\)
\item
  `Original' data-set, \(\mathcal{D}\).
\item
  (Hypothetical) \emph{reference grid data}-set, \(\mathcal{D}^{(a)}\).
\item
  Response random variable vector
  \(\mathbf{Y}=\left(Y_1,\ldots,Y_N\right)^\top\).
\item
  Observed response data vector
  \(\mathbf{y}=\left(y_1,\ldots,y_n\right)^\top\).
\item
  Generated response data vector \(\tilde{\mathbf{y}}\).
\item
  Parameter vector
  \(\boldsymbol{\theta}=\left(\theta_1,\ldots,\theta_p\right)^\top\) (of
  a model \(\mathcal{M}\)).
\item
  Probability (density or mass) function \(p\left(Y\right)\),
\item
  Conditional probability (density or mass) function
  \(p\left(Y\mid\ldots\right)\).
\item
  Observed covariate / `input' matrix \(\mathbf{X}\), that might include
  observed vectors of numerically scaled input(s), categorically scaled
  input(s), and / or grouping variable levels. In the latter case, some
  coding -- such as by using indicator functions -- needs to be
  performed in order to suitably mathematically employ those inputs in
  the model, but this would also directly be included into
  \(\mathbf{X}\). Similarly, basis function calculations as used by
  penalized spline approaches, would also just be plugged into
  \(\mathbf{X}\).
\end{itemize}

\clearpage

\section{Introduction}\label{sec-introduction}

\begin{quote}
``Bayesian modelling helps applied researchers to articulate assumptions
about their data and develop models tailored for specific applications.
Thanks to good methods for approximate posterior inference, researchers
can now easily build, use, and revise complicated Bayesian models for
large and rich data. These capabilities, however, bring into focus the
problem of model criticism. Researchers need tools to diagnose the
fitness of their models, to understand where they fall short, and to
guide their revision.'' \hfill \citep{MoranBleiRanganath2023}
\end{quote}

At the research frontiers of ecology and other academic disciplines,
researchers try to negotiate the challenge of studying complex and
dynamic systems and to thereby increase the understanding of their
fundamentals, causes and mechanisms. Such an investigated system,
accompanied by epistemic uncertainty, requires a sophisticated
scientific study approach in order to successfully explore it's
underlying latent structures. During the last decades, \emph{Bayesian
probabilistic models} -- in the following only \emph{probabilistic
model} or \emph{bayesian model} -- have emerged as an essential
statistical tool to be applied for this purpose \citep{Clark2004}. By
representing both, the parameters of the statistical model and the data,
as random variables, Bayesian models enable uncertainty updating through
the direct application of probability theory using probability
distributions \citep{SimmondsEtAl2024}. This is the main epistemic
reason for applying Bayesian modeling. However, a stronger reason in the
everyday academic work comes with progress made in computer sciences
that lead to \emph{probabilistic programming languages} such as
\emph{Stan} \citep{CarpenterEtAl2017}, allowing researchers to flexibly
customize and adapt Bayesian models to their specific needs and data
situations: incorporate hierarchical structures, latent variables, and
intricate dependencies. This makes Bayesian probabilistic programming
languages particularly well-suited for ecological data, to give just one
example. However, despite their widespread utility, the broader adoption
of Bayesian models relies on developing tools that not only make models
more adaptable, but that simplify their `routine use' for non-expert
practitioners, such as intuitive interfaces, or model-building
workflows.

The above citation quotes from an article introducing an approach for
Bayesian model checking with respect to truly observed data.
\citet{MoranBleiRanganath2023} describes an approach that allows daily
work with probabilistic models to lead to more robust insights. I will
introduce the intuitive mechanism behind the progress made by
\citet{MoranBleiRanganath2023}, and give an adaption of their approach
introducing another checking tool for, hopefully, successful application
of probabilistic models.

\emph{Model checks} are one of the most important steps in
\emph{Bayesian Workflow} \citep{GelmanEtAl2020} -- a concept that
structures the steps to be made in order to get to a trustful
probabilistic model. The existence of such a conceptual framework
already makes clear that model building is anything but trivial.
Subsequently, less suitable modeling decisions for the data at hand
\citep{MoranBleiRanganath2023}, or `buggy' computational implementation
\citep{ModrakEtAl2023}, both can lead to inaccurate inferences, which in
turn affect ecological decision-making and policy. But, mistakes are
often subtle and difficult to detect, and despite the power of Bayesian
models, many ecologists do not fully leverage modern model-building
workflow tools, such as provided by \citet{MoranBleiRanganath2023}, or
\emph{Simulation Based Calibration} check (\emph{SBC} check,
\citet{ModrakEtAl2023}), to ensure that their models are correctly
specified and implemented. SBC checks verify computational reliability
of Bayesian inference algorithms by consistency checking of posterior
with prior when the posterior is estimated on prior-predicted data. The
\emph{QOI-Check}, as introduced with this article, builds on
\citet{MoranBleiRanganath2023} and SBC in order to validate
post-estimation \emph{Quantity of Interest} (\emph{QOI}) calculation and
interpretation.

The given citation from \citet{MoranBleiRanganath2023} makes reference
to `approximate posterior inference': Statistical models are solved
numerically by using computers -- complex machines that are themselves
subject of scientific investigation. It is therefore of great importance
to check whether the models developed by researchers work as they
intended. This \emph{statistical software implementetion check} is
performed using the SBC check \citep{ModrakEtAl2023}. The SBC check is
not a workflow step that intends to check of how well a model fits
observed data -- this is what \citet{MoranBleiRanganath2023} aim for --,
but a verification tool to see if implemented models specify the
underlying truth exactly in the artificial case when the underlying
truth is indeed known \citep{ModrakEtAl2023}. In fields like software
development, rigorous practices such automated testing, or version
control, are deeply embedded, ensuring the correctness and reliability
of code. Same-goal practices are equally important in computational
modeling, where even small coding errors can have significant
consequences. Bayesian Software checking methods -- such as the SBC
Check \citep{ModrakEtAl2023} -- are therefore an important part of the
Bayesian Workflow.

In comparison to this computational checking, diagnosing the fit of an
estimated model to the observed data, in the form of residual
diagnostics for Gaussian models, or more generally by predictive checks,
is much more often conducted among applied researchers. The Bayesian
flavored \emph{Posterior Predictive Check} (PPC) assesses a model by
comparing simulated data from the posterior predictive distribution to
the observed data. In the direct definition of the PPC, the same data is
used to both fit and evaluate the model, leading to power issues of the
check. \emph{Holdout Predictive Checks} (HPCs)
\citep{MoranBleiRanganath2023} address this issue by splitting the data
into training and holdout data-sets; the model is fit to the training
data, and then the model's prediction for the holdout data is compared
to the actual holdout sample. By using separate data sets for fitting
and evaluation, HPCs provide a calibrated Bayesian check of model fit.

The message of this article is that applied researchers are not
finished, yet, when a statistical model is estimated, checked for
correct implementation, and for fit. The main remaining task is to
connect the model with the knowledge about the scientific question; the
model needs to be interpreted. This is often supported by calculation of
derived QOIs -- and their implementation and correct interpretation
needs checking, too. The QOI-Check is designed to become a step in the
\emph{Bayesian workflow} serving this purpose. Bayesian workflow
emphasizes the importance of model criticism by checking for correct
model implementation and fit. While SBC is assessing computational
validity, and Holdout Predictive Checks (HPCs) evaluate how well the
model fits the observed data, the QOI-Check focuses on the validity of
post-estimation calculations and the interpretation of quantities
derived from the model parameters.

As an example, we will see in \emph{Case Study I} a homoscedastic normal
model for response variable \(Y\),
i.e.~\(Y_{i,j}\sim\text{Normal}\left(\mu_{i,j},\sigma^2\right)\), where
a log-link is applied to the conditional expectation,
i.e.~\(\mu_{i,j} = \exp\left(\eta_{i,j}\right)\). The linear predictor,
\(\eta_{i,j}=\beta_0+\beta_1\cdot x_{i,j}+\gamma_{i}\), includes a
numeric covariate \(x_{i,j}\) and grouping coefficients -- a.k.a.
`random intercepts' --
\(\gamma_i\sim \text{Normal}\left(0,\sigma_\gamma^2\right)\) -- index
\(i=1,\ldots,G\) denotes groups, and index \(j=1,\ldots,n_i\) denotes
individual observation units. Such a model could be applied in forest
science to stem diameter increment measurements, \(Y\), of trees between
two \(5\)-yearly measurement campaigns, conditional on diameter, \(x\),
and forest stand ID, \(g\). The log link makes applied sense here since
we know that conditional on survival of the tree, a tree is expected to
have a positive growth increment in a \(5\) year long time-period.
However, \emph{measurement error} might lead to -- (rarely) occurring in
practice -- negative increment \emph{observations}, \(y<0\), which gives
the normal distribution -- distributing positive probability mass to
values below \(0\) -- it's plausibility. Interest lies here on the
diameter increment of the `average tree' at a fixed diameter,
e.g.~\(x=30~\text{cm}\), and \emph{'across stands'},
i.e.~\(\textsf{E}\left(Y\mid x=30\text{ cm}\right)\). This quantity is
not modeled directly by one single parameter, but must be calculated
post estimation using several of the model's parameters. An obvious
first guess is to use a central stand ID represented by
\(\gamma_i\overset{!}=0\) -- as advocated with the `typical farm' in a
famous statistical modeling textbook for ecologists \citep{ZuurEtAl2009}
(No offense!) --, leading to
\(\exp\left(\beta_0+\beta_1\cdot30\right)\). But by the non-linearity of
the logarithmic link-function and the symmetry of the Gaussian
distribution, \(\gamma_i=0\) will not represent the diameter increment
of the `average tree' with a diameter of \(30\) cm: given
\(\vert\gamma_i^\prime\vert=\vert\gamma_i\vert\), a positive
\(\gamma_i>0\) has a greater magnifying effect on the conditional
expectation in comparison to the reducing effect of a negative
\(\gamma_i^\prime<0\) with the same absolute value.

However, for less-complex-models such as in \emph{Case Study I},
researchers might search the literature -- which is something you have
to come up with first -- and find a calculation approach for the
\emph{conditional expectation as seen from the marginal perspective}: \[
\textsf{E}\left(Y\mid x=30\right)=\exp\left(\beta_0+\beta_1\cdot30+\frac{\sigma_\gamma^2}{2}\right),
\] as an analytical solution to: \[
\textsf{E}\left(Y\mid x = 30\right)=
\int \exp\left(\beta_0+30\cdot\beta_1+z\right)p_{\text{Normal}\left(0,\sigma_{\gamma}^2\right)}\left(Z=z\right)\,\partial z,
\] where \(p_{\text{Normal}}\left(\mu,\sigma^2\right)\) is the
probability density function of the Normal distribution, with
expectation \(\mu\), and variance \(\sigma^2\).

As a follow-up question the researcher needs to answer the question to
which underlying population definition such an `average tree
calculation' connects to? A key aspect of interpreting QOIs is
understanding the underlying population definition, since
\emph{different population definitions can lead to different
interpretations of the same quantity}. Or to put it another way, what is
the suitable population definition underlying the marginal perspective
on the conditional expectation for this statistical model? Is it `the
average of a population that is composed as the sample at hand'? Or the
average with respect to some different population definition? By using a
so denoted \emph{reference grid structure}, we can connect the QOI to a
specific (hypothetical) population definition. This connection helps us
ensure that our interpretations of the QOI are aligned with the
underlying population we are interested in.

To sum up, a QOI-Check addresses the composed question: Does the
calculation and interpretation of our QOI make sense, given a defined
population? Further, a QOI-Check can help to guide researchers in QOI
calculations for more complex models. In \emph{Case Study II}, the model
as well as the post-estimation are considerably more complex, and again
not only the question of interpretation arises, but also about
technically correct implementation. Having a means -- by the QOI-Check
-- to validate these steps leads to correct implementation and clearer
interpretation.

\subsection*{Structure of the Paper}\label{structure-of-the-paper}
\addcontentsline{toc}{subsection}{Structure of the Paper}

Section 2 describes recent approaches for Bayesian Software and Model
Checking. Section 3 describes the new approach of how to check the
implementation of post-estimation derived quantities by the use of
predictive checks on reference grids. Section 4 shows two application
examples, and Section 5 concludes with some final remarks. Additional
information is provided by supplementary materials.

\section{Bayesian Software and Model
Checking}\label{bayesian-software-and-model-checking}

\subsection*{Bayesian Inference}\label{bayesian-inference}
\addcontentsline{toc}{subsection}{Bayesian Inference}

We are given a \emph{(distributional) model}\footnote{A.k.a. the
  \emph{likelihood} or \emph{sampling model}.},
\(p\left(Y\mid\theta\right)\), for \emph{random variable}, \(Y\), where
\(\theta\) is the model's \emph{parameter} about which we want to
improve -- based on an \emph{observed data} vector
\(\mathbf{y}=\left(y_1,\ldots,y_n\right)^\top\) -- our current
(epistemic) information status. From the Bayesian perspective,
\(\theta\) itself is a random variable reflecting our current
understanding of how the parameter \(\theta\) varies within the
\emph{superpopulation}\footnote{This \emph{superpopulation viewpoint} is
  only one interpretation of distribution and variation of parameters in
  the Bayesian inference framework. \citet[Section 3.1]{Rubin1984}
  relates the prior to a superpopulation, a larger population from which
  \(\theta\) is realized. The superpopulation is a theoretical construct
  that represents an idealized population from which a sample is drawn,
  encompassing all potential observations that could arise under a
  specific model or set of conditions. The superpopulation provides a
  theoretical, epistemic framework that encompasses the broader context
  of the unknown data generating process, while the prior distribution
  reflects the researcher's uncertainty about the parameter based on
  that context. In this sense, the information represented by the prior
  distribution is shaped by the perception of the superpopulation.} --
this theroetical concept of a superpopulation only serves the purpose to
offer us a metaphorical representation of our epistemic ignorance. In
academia, we build this understanding by considering findings from
previous studies similar to ours and incorporating relevant knowledge
from other fields. Using probability as a numeric measure for our
uncertainty, a \emph{prior distribution}, \(p\left(\theta\right)\), is
formulated in order to give this understanding a numeric expression. We
are interested in the plausible values of the parameter (vector)
\(\theta\) that could have -- as an integral part of the distributional
model -- generated the observed data vector \(\mathbf{y}\). In Bayesian
inference, the distribution of plausible parameter values -- conditional
on the information increase by involvement of \(\mathbf{y}\) -- is
denoted the \emph{posterior}, \(p\left(\theta\mid\mathbf{y}\right)\).

\subsection*{Posterior Sampling}\label{posterior-sampling}
\addcontentsline{toc}{subsection}{Posterior Sampling}

Due to the lack of an analytical solution -- even if the statistical
model itself is only moderately complex --, posteriors of common
Bayesian models in nowadays research projects are approximated
numerically. Following \citet[Section 3.1]{Rubin1984} we can approximate
the posterior distribution based on a model
\(p\left(Y\mid\theta\right)\), an observed data vector \(\mathbf{y}\),
and a generative\footnote{\emph{Non-generative prior} refer to a
  probability distribution that does not provide a meaningful way to
  generate values for the purpose of simulations and / or prior
  predictive checks. `Flat' -- a.k.a. `uninformative' -- priors over an
  unbounded range, like \(p\left(\theta\right) \propto 1\) for
  \(\theta \in \mathbb{R}\) are typically non-generative since they
  don't provide a proper probability distribution for sampling.} prior
distribution \(p\left(\theta\right)\), by implementation of the
following four steps:

\begin{enumerate}
\def\labelenumi{\arabic{enumi}.}
\tightlist
\item
  Sample \(\theta_1, \ldots, \theta_S\) from the prior \(p(\theta)\).
\item
  Simulate sample-specific data-sets \(\tilde{\mathbf{y}}_s\)
  (\(s = 1, \ldots, S\)) using the model \(p(Y_i \mid \theta)\) or
  \(p(Y_i \mid \theta, \mathbf{X}_i)\), matching the dimensions of the
  observed data \(\mathbf{y}\).
\item
  Identify a subset
  \(\mathcal{S}_{\text{sub}} \subset \{1, \ldots, S\}\) where simulated
  data-sets \(\tilde{\mathbf{y}}_s\) align with \(\mathbf{y}\), with, if
  needed, using `practical rounding' for continuous data.
\item
  Retain prior samples \(\theta_s\) for
  \(s \in \mathcal{S}_{\text{sub}}\), yielding a posterior-like
  distribution of \(\theta\) consistent with \(\mathbf{y}\).
\end{enumerate}

Supplement A describes a longer version of this `toy algorithm' -- a
very naive approach to approximate the posterior distribution. But if we
keep it's `simulation spirit' in mind, it will facilitate an intuitive
understanding of the model checking approaches following later.

In order to get to a posterior approximating sample much more
computationally efficient, we rely on approximate posterior inference
methods such as \emph{Markov Chain Monte Carlo} (\emph{MCMC}), or
\emph{Hamiltonian Monte Carlo} (\emph{HMC}) \citep[Chapter 5: MCMC Using
Hamiltonian Dynamics]{Neal2011}. \emph{Probabilistic programming
languages} (\emph{PPL}s) such as \emph{Stan} \citep{CarpenterEtAl2017}
are very popular in the research community -- ecologists being no
exception \citep{MonnahanEtAl2016} -- as they allow users to efficiently
approximate posteriors for custom-build models using a high-level
programming syntax. A program written in a PPL describes the model,
prior distributions, and general structure linking from data to
parameters. \texttt{Stan} uses the \emph{No-U-Turn Sampler}
(\emph{NUTS}) \citep{HoffmanGelman2014} -- an adaptive version of HMC
that is incredible helpful for complex posterior landscapes -- for
posterior sampling. By this, \texttt{Stan} automates the process of
applying an efficient algorithm to a `model computer program', leading
to very user-friendly posterior sampling and possibly further
post-processing techniques.

In practical applications of such software, the estimation process
yields an \(S \times p\) matrix containing \(S\) samples from the
posterior distribution of a vector with \(p\) parameters. With this
matrix of posterior samples, we can then draw inferences about any
function of the parameters, known as \emph{derived quantity}
\citep{ModrakEtAl2023}, or \emph{quantity of interest} (QOI). A QOI is
often not a single parameter, but a function of multiple parameters.
This increases the chances for implementation bugs and
mis-interpretation of QOIs, making a dedicated validation method
desirable.

\subsection*{Posterior Prediction in a Simulation
Study}\label{posterior-prediction-in-a-simulation-study}
\addcontentsline{toc}{subsection}{Posterior Prediction in a Simulation
Study}

The \emph{posterior predictive distribution} of a new observation,
\(\tilde{y}\), is defined as: \[
p\left(\tilde{y}\mid y\right) =\int p\left(\tilde{y}\mid \boldsymbol{\theta}\right)p\left(\boldsymbol{\theta}\mid y\right)\partial\boldsymbol{\theta}.
\] It represents the distribution of `new' observations, given the
observed data and the fitted model, integrating over the uncertainty in
the model parameters, \(\boldsymbol{\theta}\).

In a simulation study with repeated simulation of data-sets from a
ground-truth definition for the data generating process, posterior
predictions are easily generated according to the steps of the following
simulation algorithm.

For simulation run \(r\), \(r=1,\ldots,R\):

\begin{enumerate}
\def\labelenumi{\arabic{enumi}.}
\tightlist
\item
  Draw parameter vector
  \(\boldsymbol{\theta}_{r}=\left(\theta_{1,r},\ldots,\theta_{p,r}\right)^\top\)
  from a generative prior, \(p\left(\boldsymbol{\theta}\right)\).
\item
  Use \(\boldsymbol{\theta}_{r}\) to generate the response data
  \(y_{r}\), where the distribution for generating \(y_{r}\), given
  \(\theta_{r}\), is the likelihood: \[
  y_{i,r}\sim p\left(Y \mid \boldsymbol{\theta}_{r},\mathbf{x}_i\right),\quad i=1,\ldots,n.
  \]
\item
  Once we have simulated data vector \(\mathbf{y}_{r}\) we are able to
  estimate the model's parameters posterior distribution using our
  Bayesian approximate posterior inference sampler of choice. We end up
  with \(S\) sampled parameter vectors from the posterior distribution:
  \[
  \boldsymbol{\theta}_{s,r} \sim p(\boldsymbol{\theta} \mid \mathbf{y}_{r}),\quad s=1\ldots,S
  \] that we could also organize in a \(S\times p\) posterior sample
  matrix \(\boldsymbol{\Theta}\).
\item
  For each a draw \(s\), \(s=1\ldots,S\), we select the respective
  parameter vector \(\boldsymbol{\theta}_{r,s}\) -- the \(r\)-th row in
  \(\boldsymbol{\Theta}\), and simulate -- `predict' -- a new response
  value vector \(\tilde{\mathbf{y}}_{r,s}\) from the likelihood: \[
  \tilde{y}_{r,s,i}\sim p\left(Y\mid\boldsymbol{\theta}_{r,s},\mathbf{x}_i\right),\quad i=1,\ldots,n.
  \] Using the tilde symbol, we discriminate -- by notation -- between
  predictions from the prior, \(y_i\), and predictions from the
  posterior, \(\tilde{y}_i\).
\end{enumerate}

In the 4\textsuperscript{th} step, we can think of predicting for
different reference data-sets. Of course, we can use the original
covariate data structure \(\mathcal{D}\) that is attached to \(y_r\), or
we can generate a new reference grid, \(\mathcal{D}^{(a)}\), introducing
a different covariate structure.

For an observation unit \(i\), \(i=1,\ldots,N\) that belongs to the
original data, \(\mathcal{D}\): \[
\tilde{y}_{r,s,i} \mid \boldsymbol{\theta}_{r,s} \sim p\left(Y \mid \boldsymbol{\theta}_{r,s},\mathbf{X}_i\right).
\]

For an observation unit \(j\), \(j\notin\left\{1,\ldots,N\right\}\),
that belongs to some `new' holdout-data, or reference grid data,
\(\mathcal{D}^{(a)}\): \[
\tilde{y}_{r,s,j} \mid \boldsymbol{\theta}_{r,s} \sim p\left(Y \mid \boldsymbol{\theta}_{r,s},\mathbf{X}_j\right).
\]

\subsection*{Posterior Predictive
Checks}\label{posterior-predictive-checks}
\addcontentsline{toc}{subsection}{Posterior Predictive Checks}

A \emph{Posterior Predictive Check} (\emph{PPC}) is a Bayesian model
evaluation technique that assesses the fit of a model by comparing
observed data to data generated from the posterior predictive
distribution. The comparison in PPCs is often performed visually
\citep{GabryEtAl2019}, such as by using kernel density methods, or the
empirical cumulative density function. Alternatively, not the full
distribution, but on or several specific attributes -- such as mean,
selected quantiles, standard deviation, \ldots -- of the distribution
are compared, using a diagnostic function or discrepancy measure,
\(T\left(y\right)\), or \(T\left(y,\boldsymbol{\theta}\right)\), that
summarizes the respective attribute.

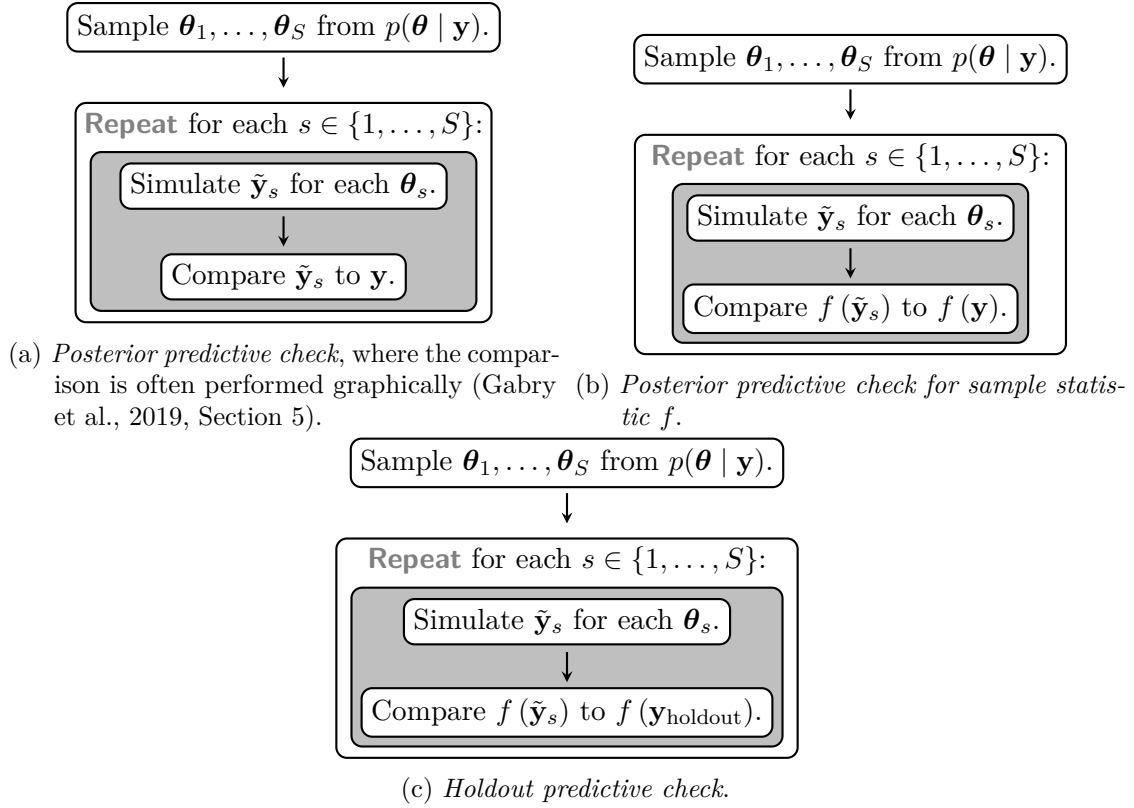
\begin{figure}[!h]
\centering
\begin{subfigure}{0.49\textwidth}
\centering
\begin{tikzpicture}[node distance=1.5cm, thick, >=stealth, rounded corners]
    \tikzstyle{box} = [draw, minimum width=5.1cm, minimum height=2.1cm]
    \node[box, fill = lightgray] at (0, .4) {};
    \node[rectangle, draw, text centered, minimum height=0cm, minimum width=3cm] at (0, 3.1) {Sample \( \boldsymbol{\theta}_1, \dots, \boldsymbol{\theta}_S \) from \( p(\boldsymbol{\theta} \mid \mathbf{y}) \).};
    \node at (0, 1.8) {\textcolor{gray}{\textbf{\textsf{Repeat}}} for each $s\in\left\{1,\ldots,S\right\}$:};
    \node[rectangle, draw, fill = white, text centered, minimum height=0cm, minimum width=3cm] at (0, 1) 
        {Simulate \( \tilde{\mathbf{y}}_s \) for each \( \boldsymbol{\theta}_s \).};
    \node[rectangle, draw, fill = white, text centered, minimum height=0cm, minimum width=3cm] at (0, -.2) 
        {Compare \( \tilde{\mathbf{y}}_s \) to \( \mathbf{y} \).};
    \draw[->] (0, 2.7) -- (0, 2.3);
    \draw[->] (0, 0.6) -- (0, .2);
    \tikzstyle{box} = [draw, minimum width=5.5cm, minimum height=2.9cm]
    \node[box] at (0, .65) {};
\end{tikzpicture}
\caption{\emph{Posterior predictive check}, where the comparison is often performed graphically \citep[Section 5]{GabryEtAl2019}.}
\end{subfigure}
\hfill
\begin{subfigure}{0.49\textwidth}
\centering
\begin{tikzpicture}[node distance=1.5cm, thick, >=stealth, rounded corners]
    \tikzstyle{box} = [draw, minimum width=4.7cm, minimum height=2.1cm]
    \node[box, fill = lightgray] at (0, .4) {};
    \node[rectangle, draw, text centered, minimum height=0cm, minimum width=3cm] at (0, 3.1) {Sample \( \boldsymbol{\theta}_1, \dots, \boldsymbol{\theta}_S \) from \( p(\boldsymbol{\theta} \mid \mathbf{y}) \).};
    \node at (0, 1.8) {\textcolor{gray}{\textbf{\textsf{Repeat}}} for each $s\in\left\{1,\ldots,S\right\}$:};
    \node[rectangle, draw, fill = white, text centered, minimum height=0cm, minimum width=3cm] at (0, 1) 
        {Simulate \( \tilde{\mathbf{y}}_s \) for each \( \boldsymbol{\theta}_s \).};
    \node[rectangle, draw, fill = white, text centered, minimum height=0cm, minimum width=3cm] at (0, -.2) 
        {Compare \(f\left(\tilde{\mathbf{y}}_s\right)\) to \(f\left(\mathbf{y}\right) \).};
    \draw[->] (0, 2.7) -- (0, 2.3);
    \draw[->] (0, 0.6) -- (0, .2);
    \tikzstyle{box} = [draw, minimum width=5.7cm, minimum height=2.9cm]
    \node[box] at (0, .65) {};
\end{tikzpicture}
\caption{\emph{Posterior predictive check for sample statistic} $f$.}
\end{subfigure}
\begin{subfigure}{0.49\textwidth}
\centering
\begin{tikzpicture}[node distance=1.5cm, thick, >=stealth, rounded corners]
    \tikzstyle{box} = [draw, minimum width=5.7cm, minimum height=2.1cm]
    \node[box, fill = lightgray] at (0, .4) {};
    \node[rectangle, draw, text centered, minimum height=0cm, minimum width=3cm] at (0, 3.1) {Sample \( \boldsymbol{\theta}_1, \dots, \boldsymbol{\theta}_S \) from \( p(\boldsymbol{\theta} \mid \mathbf{y}) \).};
    \node at (0, 1.8) {\textcolor{gray}{\textbf{\textsf{Repeat}}} for each $s\in\left\{1,\ldots,S\right\}$:};
    \node[rectangle, draw, fill = white, text centered, minimum height=0cm, minimum width=3cm] at (0, 1) 
        {Simulate \( \tilde{\mathbf{y}}_s \) for each \( \boldsymbol{\theta}_s \).};
    \node[rectangle, draw, fill = white, text centered, minimum height=0cm, minimum width=3cm] at (0, -.2) 
        {Compare \(f\left(\tilde{\mathbf{y}}_s\right)\) to \(f\left(\mathbf{y}_{\text{holdout}}\right) \).};
    \draw[->] (0, 2.7) -- (0, 2.3);
    \draw[->] (0, 0.6) -- (0, .2);
    \tikzstyle{box} = [draw, minimum width=6.1cm, minimum height=2.9cm]
    \node[box] at (0, .65) {};
\end{tikzpicture}
\caption{\emph{Holdout predictive check}.}
\end{subfigure}
\caption{Illustration of \emph{posterior predictive checks}.}
\label{fig-posterior_holdout_checks}
\end{figure}

Figure \ref{fig-posterior_holdout_checks} on page
\pageref{fig-posterior_holdout_checks} shows a sketch for posterior
predictive checking -- following \citet{GabryEtAl2019}.

\begin{quote}
``If the model fits, then replicated data generated under the model
should look similar to observed data. To put it another way, the
observed data should look plausible under the posterior predictive
distribution. This is really a self-consistency check: an observed
discrepancy can be due to model misfit or chance. Our basic technique
for checking the fit of a model to data is to draw simulated values from
the joint posterior predictive distribution of replicated data and
compare these samples to the observed data. Any systematic differences
between the simulations and the data indicate potential failings of the
model.'' \hfill \citep{GelmanEtAl2013}
\end{quote}

PPCs use the same data \added{twice, once} to fit the model, and then a
second time to evaluate the model fit. This leads to overconfident
assessments of the quality of a model, and therefore an improperly
calibrated check \citep{MoranBleiRanganath2023}. A \emph{posterior
predictive p-value} \citep{GelmanEtAl2013} is defined as the probability
that the replicated data could be more extreme than the observed data,
as measured by \(T\left(y\right)\), or
\(T\left(y,\boldsymbol{\theta}\right)\): \[
p_{\text{posterior predictive}} = p\left(T\left(\tilde{y}, \boldsymbol{\theta}\right) \geq T\left(y, \boldsymbol{\theta}\right)\mid y\right)
\] The symptom of improper calibration is that the
\(p_{\text{posterior predictive}}\) is non-uniformly distributed, with a
concentration around \(0.5\) \citep{MoranBleiRanganath2023}.

\subsection*{Holdout Predictive Check and Split Predictive
Check}\label{holdout-predictive-check-and-split-predictive-check}
\addcontentsline{toc}{subsection}{Holdout Predictive Check and Split
Predictive Check}

A \emph{Holdout Predictive Check} \citep{MoranBleiRanganath2023}, short
\emph{HPC}, evaluates the fit of a Bayesian model by testing whether the
model can accurately predict \emph{held-out data}. HPCs split the data
into training and holdout sets, fit the model to the training data, and
then compare predictions for the holdout data to the actual values. This
avoids the `double use of data' as in PPCs. HPCs are properly calibrated
for asymptotically normal diagnostics and have been empirically shown to
be calibrated for other types of diagnostics
\citep{MoranBleiRanganath2023}.

\begin{quote}
``A well-calibrated method for model checking must correctly reject a
model that does not capture aspects of the data viewed as important by
the modeler and fail to reject a model fit to well-specified data.''
\hfill \citep{LiHuggins2024}
\end{quote}

\subsection*{Prior Predictive Check in a Simulation
Study}\label{prior-predictive-check-in-a-simulation-study}
\addcontentsline{toc}{subsection}{Prior Predictive Check in a Simulation
Study}

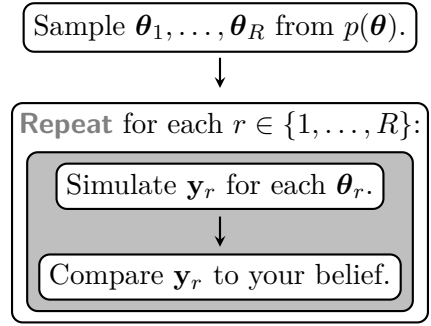
\begin{figure}[!h]
\centering
\begin{tikzpicture}[node distance=1cm, thick, >=stealth, rounded corners]
    \tikzstyle{box} = [draw, minimum width=5.1cm, minimum height=2.1cm]
    \node[box, fill = lightgray] at (0, .4) {};
    \node[rectangle, draw, text centered, minimum height=0cm, minimum width=3cm] at (0, 3.1) {Sample \( \boldsymbol{\theta}_1, \dots, \boldsymbol{\theta}_R \) from \( p(\boldsymbol{\theta}) \).};
    \node at (0, 1.8) {\textcolor{gray}{\textbf{\textsf{Repeat}}} for each $r\in\left\{1,\ldots,R\right\}$:};
    \node[rectangle, draw, fill = white, text centered, minimum height=0cm, minimum width=3cm] at (0, 1) 
        {Simulate \( \mathbf{y}_r \) for each \( \boldsymbol{\theta}_r \).};
    \node[rectangle, draw, fill = white, text centered, minimum height=0cm, minimum width=3cm] at (0, -.2) 
        {Compare \( \mathbf{y}_r \) to your belief.};
    \draw[->] (0, 2.7) -- (0, 2.3);
    \draw[->] (0, 0.6) -- (0, .2);
    \tikzstyle{box} = [draw, minimum width=5.5cm, minimum height=2.9cm]
    \node[box] at (0, .65) {};
\end{tikzpicture}
\caption{\emph{Prior Predictive Check}.}
\label{fig:prior_predictive_check}
\end{figure}

A \emph{Prior Predictive Check} is a Bayesian Workflow element that
allows researchers to see how well their domain expertise aligns with
the prior information stored in the parameter specific prior
distributions. The Prior Predictive Check is technically very similar to
the PPC, with the only difference is that the influence of the data
\(y\) in the form of the likelihood function is ignored in the Prior
Predictive Check. This check is usually performed visually -- as for the
PPC, and again sometimes augmented by summary statistics of the prior
predicted distribution. Figure \ref{fig:prior_predictive_check} on page
\pageref{fig:prior_predictive_check} shows a sketch for prior predictive
checking -- following \citet[Section 3]{GabryEtAl2019}.

\subsection*{Simulation Based
Calibration}\label{simulation-based-calibration}
\addcontentsline{toc}{subsection}{Simulation Based Calibration}

\emph{Simulation-Based Calibration (SBC)} \citep{ModrakEtAl2023}, in
particular, has emerged as a powerful tool for verifying the
implementation of probabilistic models by checking whether the model's
posterior distribution behaves as expected when data are simulated from
the model itself. Subfigure (a) in Figure
\ref{fig-simulation_based_calibration} on page
\pageref{fig-simulation_based_calibration} gives an illustration of SBC
following \citet{TaltsEtAl2018}, Subfigure (b) presents an illustration
of SBC following \citet{ModrakEtAl2023}. In short, the SBC check is
build by the following steps:

\begin{description}
\item[{[}SBC Check `algorithm'{]}]
Repeat \(R\) times: (prior sampling \(\rightarrow\) data simulation
\(\rightarrow\) posterior estimation, i.e.~posterior sampling
\(\rightarrow\) calculate parameter-wise ranking of prior sample among
posterior samples).
\end{description}

The SBC check starts with drawing samples from the prior -- similar to
the above description of the algorithmic superpopulation viewpoint on
Bayesian inference by \citet{Rubin1984}. Based on a vector of simulated
parameters -- one draw for each of the model's parameters -- a data-set
\(\mathbf{y}_r\) is simulated -- in regression usually for the given
covariate(s) observations denoted \(\mathbf{X}\). Based on
\(\mathbf{y}_{r}\) a posterior sample is generated, for example by using
\emph{Stan} \citep{CarpenterEtAl2017}. This posterior sample is based on
the exactly similar likelihood model that was used to generate the data,
and on the exactly similar prior that was used to draw the prior sample.
Post estimation, the rank of the prior draw within the posterior draws
is calculated for each of the parameters. This parameter-wise
calculation of the rank of a prior sample within the distribution of
posterior samples is performed by treating the prior sample as part of
the set of posterior samples, ranking all values in ascending order
(e.g., analogous to a competition where lower values rank higher), and
identifying the rank position of the prior sample.

For example, if the prior sample value is \(\theta_0=0.32\) and the
posterior samples are \(\left\{0.1,0.2,0.4\right\}\), the set becomes
\(\left\{0.1,0.2,0.32,0.4\right\}\), and \(\theta_0\) occupies the third
position in this ranking.

The estimation is computationally trustworthy, i.e.~it passes the SBC
check, if the collected -- over \(R\) runs -- ranking positions are
uniformly distributed, which can be checked when this algorithm is
repeated several, \(R\), times \citep{ModrakEtAl2023}.

\subsection*{Checking QOIs and the Population They
Represent}\label{checking-qois-and-the-population-they-represent}
\addcontentsline{toc}{subsection}{Checking QOIs and the Population They
Represent}

Most statistical regression models that are not based on the
\emph{Gaussian} or \emph{t} distribution connect the conditional
expectation to the linear predictor through a non-linear link function.
Hence, regression parameters -- modeling changes in the linear predictor
for the conditional expectation -- operate on a transformed scale,
rather than on the original scale of the response variable such as in a
model using the identity link function. Consequently, if one wishes to
interpret parameters of such kind on the original response scale,
inverse link functions must be applied, leading to -- in most
applications -- consideration of more than the current focus parameter.
So interpretation of the QOI
\emph{'conditional expectation on the original scale of the response variable'},
in such a model with a non-linear link function, requires calculation of
a quantity that is derived out of several model parameters. The
QOI-Check can confirm that the QOI is computed correctly and further,
that is interpreted appropriately within the context of a chosen
population, as can be seen in the following example.

As the example in \emph{Case Study I}, I will use a Gaussian response
model: \[
y_{ij}\sim\text{Normal}\left(\mu_{ij},\sigma^2\right),
\] with log-link for the conditional expectation
\(\text{E}\left(y_{ij}\mid\mathbf{X}_{ij}\right)=\mu_{ij}\): \[
\mu_{ij} = \exp\left(\eta_{ij}\right),
\] where the linear predictor is a function of input data
\(\mathbf{X}_{ij}\), for example
\(\eta_{ij}=\beta_0+\beta_1\cdot x_{ij}\).

When a multilevel model, for example for repeated observations, index
\(j\), within groups, index \(i\), uses such a non-identity link
function, the interpretation of the conditional expectation, as our QOI
in \emph{Case Study I}, further differs from that in a standard Gaussian
multilevel model with an identity link function:

\[
\text{E}\left(y_{ij}\mid\mathbf{X}_{ij}\right)=\exp\left(\beta_0+\beta_1\cdot x_{ij}\right)\exp\left(\gamma_{i}\right),
\] where \(\gamma_i\) denote group-specific deviations from `an average'
on the scale of the linear predictor. All estimates for parameters
\(\beta_0\), \(\beta_1\), and \(\gamma_i\), in this model have their
effects on the log scale, meaning they lack direct interpretation on the
response scale.

It is essential to recall here that effects in (generalized) linear
multilevel models have cluster-specific interpretations -- often called
`subject-specific' in longitudinal models when clusters represent
subjects. Researchers accustomed to working with linear multilevel
models may sometimes overlook this distinction, as in linear multilevel
models, cluster-specific effects are equivalent to population-average
(marginal) effects. However, in (generalized) linear multilevel models,
a non-linear link-function means that reversing the link such as the log
link in this example, does not yield expected values on the natural
scale as a consequence of the distribution of random effects and
Jensen's inequality.

In the marginal perspective on such a model, we are interested in the
distribution of the response variable in a hypothetical scenario where
we approach a new, `unseen', group and take our first observation there.
In our \emph{Case Study I} scenario, with the linear mixed model with
one random intercept term and log-link function, we don't see marginal
effects directly, as in linear multilevel models. However, it is known
that adding half the grouping-coefficient's distributions
variance\footnote{It is not the grouping-coefficients' variance, but the
  variance of the distribution underlying the grouping-coefficients.} to
the `fixed effects' linear predictor, \(\beta_0+\beta_1\cdot x_{ij}\)
allows us to move from the subject-specific to the marginal perspective
for model with logarithmic link-function.

This broader interpretation involves Quantities of Interest -- or
\emph{generated quantities} -- that combine multiple parameters from the
estimated model, requiring correct specification for meaningful results.
Additionally, applied researchers may find it challenging to identify
the appropriate population definition for accurately interpreting these
averaged effects. Is the proper population definition to
\(\exp\left(\beta_0+\beta_1\cdot x_{ij}+\frac{\sigma_\gamma^2}{2}\right)\)
definition the one that is of similar group composition as in the data
sample, or is a balanced grouping structure the correct definition? The
QOI-Check is able to select to correct population definition among
several alternatives.

\section{Prior-Posterior Derived-Predictive Consistency Checks
(`QOI-Check')}\label{prior-posterior-derived-predictive-consistency-checks-qoi-check}

The \emph{QOI-Check} is specified from two different viewpoints, which
are the \emph{Prior-Derived Posterior-Predictive Consistency Check}, and
the \emph{Prior-Predicted Prior-Derived Consistency Check}. Both
versions of the OOI-Check are designed to ensure the validity of
post-estimation QOIs. The \emph{Prior-Derived Posterior-Predictive}
version is particularly useful when the applied focus of the QOI is a
feature of the predictive distribution for a specific definition of the
population of new observation units. In contrast, the
\emph{Prior-Predicted Posterior-Derived} version is more natural when we
want to assess the correctness of QOI post-estimation (software)
implementation, i.e.~the QOI being more closely related to a model
inference statement for one new observation unit than to a population
feature. However, in practice this distinction seems to play only a
minor role as long as the QOI-Check is carried out in any of the two
ways. \emph{Case Study I} applies both version interchangeably,
\emph{Case Study II} only uses the \emph{Prior-Predicted
Posterior-Derived} version since it is here of particular applied
workflow interest that the QOI is correctly implemented -- at a new
observation unit -- post estimation.

As in SBC and HPC, each of the two QOI-check version leads to inequality
statements between one prior quantity and many posterior quantities,
leading to an uniformity check that judges the consistency between the
prior and the posterior utilization. Here, the QOI-Check helps to
validate QOI calculations by different population definitions
implemented through \emph{data structures}, making it clear to
researchers exactly what population they are making claims about.

\subsubsection*{Data structures}\label{data-structures}
\addcontentsline{toc}{subsubsection}{Data structures}

Prior and posterior predictions may depend on different data structures,
\(\mathcal{D}_a\), and \(\mathcal{D}_b\), respectively. These data
structures are crucial components of the QOI-Check.

The \emph{replicate simulation-data structure} re-uses the same
covariate structure as given for some original covariate and grouping
composition of the sample, and employs it on parameter values that are
generated, ie. `sampled', from the prior, or posterior.

In contrast, a \emph{reference grid structure} contains a structural
change to the original covariate and grouping composition of the sample.
For example in the case of a model with a grouping variable, a reference
grid might include new levels of this variable, representing unseen
compartments of the underlying population. Such a reference grid
structure applies the model's ability to generalize beyond the observed
data, and they allow us to evaluate the QOI under different covariate
structures than those used for model fitting. Reference grid structures
might make it necessary to generate new model parameters, such as for
levels of a grouping variable that were not included in the original
data, \(\mathcal{D}\). This leads to the potentially extended model
parameters vector
\(\boldsymbol{\theta}_{r}\left[,\boldsymbol{\theta}_{r}^{\left(\mathcal{D}_a\right)}\right]\)
as used in the following inequality equations defining the QOI-Check
versions.

\subsection*{Prior-Derived Posterior-Predictive Consistency
Check}\label{prior-derived-posterior-predictive-consistency-check}
\addcontentsline{toc}{subsection}{Prior-Derived Posterior-Predictive
Consistency Check}

This first version of the QOI-Check is based on the following inequality
statement: \[
f\left(\boldsymbol{\theta}_{r}\left[,\boldsymbol{\theta}_{r}^{\left(\mathcal{D}_a\right)}\right]\right)
<
g\left(\tilde{\mathbf{y}}_{r,s}^{\left(\mathcal{D}_b\right)}\right).
\] Function \(f\) returns the QOI as a direct function of the
potentially extended model parameters -- therefore denoted \emph{`prior
derived'} --, and function \(g\) returns the QOI as a sample statistic
-- such as the arithmetic mean, for example -- of posterior predicted
values. The posterior predicted values might potentially also depend on
an extended model parameters vector
\(\boldsymbol{\theta}_{r}\left[,\boldsymbol{\theta}_{r}^{\left(\mathcal{D}_b\right)}\right]\).

This inequality is evaluated for each of the posterior samples, \(s\),
and everything is repeated for the \(R\) simulation runs,
\(r=1,\ldots,R\).

Figure \ref{fig:Prior-Derived-Posterior-Predictive} on page
\pageref{fig:Prior-Derived-Posterior-Predictive} shows a sketch of this
version of the QOI-Check.

\subsection*{Prior-Predicted Posterior-Derived Consistency
Check}\label{prior-predicted-posterior-derived-consistency-check}
\addcontentsline{toc}{subsection}{Prior-Predicted Posterior-Derived
Consistency Check}

This second version of the QOI-Check is based on the following
inequality equation: \[
f\left(\tilde{\mathbf{y}}_{r}^{\left(\mathcal{D}_a\right)}\right)
        <
        g\left(\boldsymbol{\theta}_{r,s}\left[,\boldsymbol{\theta}_{r,s}^{\left(\mathcal{D}_b\right)}\right]\right)
\] Function \(f\) returns the QOI as a sample statistic -- such as the
arithmetic mean, for example -- of prior predicted values of the
response variable \(y\) for data structure \(\mathcal{D}_a\) --
therefore denoted \emph{`prior predicted'} --, and function \(g\)
returns the QOI as a direct function of the potentially extended model
parameters' posterior sample. The prior predicted values might here also
potentially depend on an extended model parameters vector
\(\boldsymbol{\theta}_{r}\left[,\boldsymbol{\theta}_{r}^{\left(\mathcal{D}_a\right)}\right]\).

By the same reasoning as for the first definition of a QOI-Check
version, the `potentially extended vector notation',
\(\boldsymbol{\theta}_{r,s}\left[,\boldsymbol{\theta}_{r,s}^{\left(\mathcal{D}_b\right)}\right]\),
is applied again.

Figure \ref{fig:Prior-Predictive-Posterior-Derived} on page
\pageref{fig:Prior-Predictive-Posterior-Derived} shows a sketch of the
second variant of the proposed approach.

\subsection*{Software}\label{software}
\addcontentsline{toc}{subsection}{Software}

The QOI-Check as implemented for \emph{Case Studies I and II} is
implemented in the statistical software environment \emph{R}
\citep{RCoreTeam2024}, relies heavily on the R add-on package \emph{SBC}
\citep{ModrakEtAl2023}, and it's internal connection to the \emph{brms}
package \citep{Buerkner2017, Buerkner2018}, which builds upon the
probabilistic programming language \emph{Stan} \citep{CarpenterEtAl2017}
for model implementation and posterior sampling.

Further, R add-on packages \emph{ggplot2} \citep{Wickham2016},
\emph{colorspace} \citep{StaufferEtAl2009}, and \emph{cowplot}
\citep{Wilke2023} are used for graphical visualizations, and R add-on
package \emph{plyr} \citep{Wickham2011} for general data-management.

\section{Case Studies}\label{case-studies}

\begin{description}
\item[Simulation and sampling indexing]
Simulation runs are indexed by \(r=1,\ldots,R\), and sampling index,
\(s\), denotes a posterior sample,
\(s=s_\text{posterior}\in\left\{1,\ldots,S\right\}\).
\end{description}

\subsection*{Case Study I: Marginal expectation in a multilevel model
with log
link}\label{case-study-i-marginal-expectation-in-a-multilevel-model-with-log-link}
\addcontentsline{toc}{subsection}{Case Study I: Marginal expectation in
a multilevel model with log link}

This first case study uses and compares \emph{prior-derived
posterior-predictive} and \emph{prior-predictive posterior-derived
consistency checks} for the \emph{conditional expectation from the
marginal perspective} as the \emph{quantity of interest} in the
homoscedastic multilevel normal model with log-link -- as already
described in Section~\ref{sec-introduction}, and in more detail the
coming subsections.

\begin{description}
\item[Likelihood Model]
We base the simulation for this case study on the homoscedastic normal
distribution for the response, \(Y_{ij}\), with a log-link applied to
the conditional expectation, \(\mu_{ij}\). The linear predictor,
\(\eta_{ij}\), includes a numeric covariate \(x_{ij}\) and grouping
coefficients\footnote{A.k.a. `random intercepts'.}, \(\gamma_j\): \[
Y_{ij}\sim\text{Normal}\left(\mu_{ij},\sigma^2\right),\,\,\mu_{ij}=\exp\left(\eta_{ij}\right),\,\,\eta_{ij}=\beta_0+\beta_1\cdot x_{ij}+\gamma_{i},\,\,\gamma_i\sim \text{Normal}\left(0,\sigma_\gamma^2\right).
\]
\end{description}

Index \(i=1,\ldots,G\) marks grouping in \(G\) groups, and index
\(j=1,\ldots,n_i\) labels the individual observation unit index within
groups, leading to a simulated data-set of \(N=\sum_{i}n_i\) rows in
total. An indicator function\footnote{Implementing \emph{dummy coding},
  i.e.~\(\mathrm{I}_{\left\{\text{condition}\right\}}=1\) if the
  condition is met, and
  \(\mathrm{I}_{\left\{\text{condition}\right\}}=0\) otherwise.},
\(\mathrm{I}_{\left\{\text{condition}\right\}}\), is used in the
software implementation to link the elements, \(g_{i}\), of the grouping
variable observation vector to the grouping coefficients,
\(\gamma_{ij}\).
i.e.~\(\eta_{ij}=\beta_0+\beta_1 x_{ij}+\sum_{k=1}^{G} \mathrm{I}_{\left\{g_{ij}=k\right\}}\gamma_{k}\).

\begin{description}
\item[Covariate Structure \(\mathcal{D}\) (as used during Posterior
Sampling)]
Each of the \(G\) groups is to be included into a simulated data-set by
random sampling with replacement on equal probabilities, leading to the
grouping variable observation vector,
\(\left(g_1,\ldots,g_N\right)^\top\). The observation vector
\(\left(x_1,\ldots,x_N\right)^\top\) of continuous covariate \(x\) is
simulated using the continuous uniform distribution between \(0\) and
\(2\), i.e.~\(X\sim\text{Uniform}\left[0,2\right]\).
\end{description}

\vspace{0mm}

\begin{description}
\item[Prior Specification]
We set the following prior distribution specifications:
\(\beta_{0}\sim\text{Normal}\left(0, 0.1^2\right)\),
\(\beta_{1}\sim\text{Normal}\left(1, 0.1^2\right)\),
\(\sigma_\gamma \sim \text{Normal}_{+}\left(0.5, 0.1^2\right)\),
\(\sigma \sim \text{Normal}_{+}\left(1, 0.1^2\right)\), where
\(\text{Normal}_{+}\) denotes the half-normal distribution truncated to
only positive real numbers.
\end{description}

\vspace{0mm}

\begin{description}
\item[Quantity of Interest: \emph{Marginal Perspective Conditional
Expectation}]
As described in Section~\ref{sec-introduction}, we are interested in the
\emph{marginal expectation at a fixed} \(x\), here in this simulation
\(x=1\). This quantity isn't directly represented by a single parameter;
instead, it must be calculated after estimation using multiple
parameters from the model.
\end{description}

\vspace{0mm}

\begin{description}
\item[Conditional Perspective Conditional Expectation Equation
{[}Version `(a)'{]}]
This 1\textsuperscript{st} version for the conditional expectation
equation omits half grouping variable coefficients variance,
\(\frac{\sigma^2_{\gamma,s}}{2}\): \[
\textsf{E}_{(a)}\left(Y\mid x=1\right)=\exp\left(\beta_{0,s}+\beta_{1,s}\right).
\] For example, this \emph{conditional perspective} is taken in
\citet{ZuurEtAl2009} in an analyses of infection proportions of Deer
grouped in several farms. \citet{ZuurEtAl2009} denote this
`\(\gamma_i\overset{!}{=}0\) perspective' the `typical farm' scenario.
We use this 1\textsuperscript{st} version as a benchmark to see if and
how the conditional perspective relates to the marginal perspective in
this likelihood model of Case Study I.
\end{description}

\vspace{0mm}

\begin{description}
\item[Marginal Perspective Conditional Expectation Equation {[}Version
`(b)'{]}]
This 2\textsuperscript{nd} version for the conditional expectation
equation includes half the grouping variable coefficients variance,
\(\frac{\sigma^2_{\gamma,s}}{2}\): \[
\textsf{E}_{(b)}\left(Y\mid x=1\right)=\exp\left(\beta_{0,s}+\beta_{1,s}+\frac{\sigma^2_{\gamma,s}}{2}\right).
\]
\end{description}

\vspace{0mm}

\begin{description}
\item[Prior / Posterior Data-Structure]
We set up two data-structures for generation of prior and posterior
predictions:
\end{description}

\begin{itemize}
\item
  \emph{Replicate simulation-data structure} \(\mathcal{D}_{a}\): The
  simulation structure for predictor variable \(g\) is re-used, but for
  each simulation iteration \(r\), new grouping coefficients
  \(\left(\theta_{G+1},\ldots,\theta_{2\cdot G}\right)^\top_{s_{\text{prior}},r}\)
  are generated using \(\sigma_{\theta,s_{\text{prior}},r}\). This leads
  to \(\tilde{y}_{(a),i}\), \(i=1,\ldots,N_{(a)}\) in order to get to
  new simulated outcome values, \(\tilde{y}_{(A),i}\),
  \(i=1,\ldots,N_{(A)}\), from the (prior or posterior) predictive
  distribution.
\item
  \emph{Reference grid structure} \(\mathcal{D}_{b}\): We generate
  \(G=200\) new equally grouping variable levels with exactly one
  observation for each. In order for this to be applied, we have to come
  up with a definition for grouping coefficients
  \(\left(\theta_{21},\ldots,\theta_{220}\right)^\top_{s_{\text{prior}}}\)
  using \(\sigma_{\theta,s_{\text{prior}}}\). This leads to
  \(\tilde{y}_{(B),k}\), \(k=1,\ldots,N_{(B)}\).
\end{itemize}

Figure \ref{fig-tikz--A} on page \pageref{fig-tikz--A} shows a sketch --
inspired by \citet[Figure 1]{Little1993} -- for the prediction
data-structures applied in this case study. Subfigure (a) illustrates
the \emph{replicate structure}, i.e.~the same structure that was
foundational to the simulation of the observed response data vector
\(\mathbf{y}\). Subfigure (b) illustrates the \emph{reference grid} used
in this application, which is most appropriately described as `many new
minimally small groups'.

\textsf{\textbf{Sample mean [Version (c)]:}} For prior or posterior samples, $s$, we can generate simulations from the replicate simulation-data structure or the reference grid, and calculate their respective arithmetic sample means:

\[
\textsf{E}_{(c,A),s}\left(Y\mid x=1\right)=\dfrac{1}{N_{(A)}}\sum\limits_{k=1}^{N_{(A)}}\tilde{y}_{(A),k,s},
\qquad
\textsf{E}_{(c,B),s}\left(Y\mid x=1\right)=\dfrac{1}{N_{(B)}}\sum\limits_{k=1}^{N_{(B)}}\tilde{y}_{(B),k,s}.
\]

\vspace{0mm}

\begin{description}
\item[Prior or Posterior Prediction with Grouping Coefficients
\(\gamma_i\)]
When generating predictions including new levels of a grouping variable,
we basically have two options:
\end{description}

\begin{itemize}
\tightlist
\item
  With \emph{uncertainty sampling}, we re-use the generated posterior
  samples, \(\gamma_{i,s}\), of the grouping variable coefficients. This
  strategy is implemented by
  \texttt{sample\_new\_levels\ =\ \textquotesingle{}uncertainty\textquotesingle{}}
  in \texttt{brms::prepare\_predictions}
  \citep{Buerkner2017, Buerkner2018}.
\item
  With \emph{Gaussian sampling}, we generate new coefficients by
  post-estimation sampling from the `shared distribution'
  \(\gamma_i\sim\text{Normal}\left(0,\sigma_{\gamma,s}^2\right)\). This
  strategy is implemented by
  \texttt{sample\_new\_levels\ =\ \textquotesingle{}gaussian\textquotesingle{}}
  in \texttt{brms::prepare\_predictions}
  \citep{Buerkner2017, Buerkner2018}.
\end{itemize}

\vspace{0mm}

\begin{description}
\item[Uniformity Check]
We compare each of the following prior sample applications to each of
the posterior sample applications (results are visualized in Figure
\ref{fig-log-link-multilevel-model--C} on page
\pageref{fig-log-link-multilevel-model--C}, position statements for the
Subfigure rows and columns are given in brackets in the following
lists):
\end{description}

(1\textsuperscript{st} row) \(\textsf{E}_{(a),s_\text{prior}}\): Version
(a) with using \(\beta_{0,s_\text{prior}}\) and
\(\beta_{1,s_\text{prior}}\),

(2\textsuperscript{nd} row) \(\textsf{E}_{(b),s_\text{prior}}\): Version
(b) with using \(\beta_{0,s_\text{prior}}\),
\(\beta_{1,s_\text{prior}}\) and \(\sigma_{\gamma,s_\text{prior}}\),

(3\textsuperscript{rd} row) \(\textsf{E}_{(c,A,G),s_\text{prior}}\):
Sample mean of prior prediction on reference grid structure with
Gaussian sampling for \(\gamma_{i,s_\text{prior}}\) from
\(\gamma_{i,s_\text{prior}}\sim\text{Normal}\left(0,\sigma_{\gamma,s_\text{prior}}^2\right)\),

(4\textsuperscript{th} row) \(\textsf{E}_{(c,A,u),s_\text{prior}}\):
Sample mean of prior prediction on reference grid structure with
uncertainty sampling for \(\gamma_{i,s_\text{prior}}\),

(5\textsuperscript{th} row) \(\textsf{E}_{(c,B,G),s_\text{prior}}\):
Sample mean of prior prediction on replicate simulation-data structure
with Gaussian sampling for \(\gamma_{i,s_\text{prior}}\) from
\(\gamma_{i,s_\text{prior}}\sim\text{Normal}\left(0,\sigma_{\gamma,s_\text{prior}}^2\right)\),

(6\textsuperscript{th} row) \(\textsf{E}_{(c,B,u),s_\text{prior}}\):
Sample mean of prior prediction on replicate simulation-data with
uncertainty sampling for \(\gamma_{i,s_\text{prior}}\),

vs.

(1\textsuperscript{st} column) \(\textsf{E}_{(a),s_\text{posterior}}\):
Version (a) with using \(\beta_{0,s_\text{posterior}}\) and
\(\beta_{1,s_\text{posterior}}\),

(2\textsuperscript{nd} column) \(\textsf{E}_{(b),s_\text{posterior}}\):
Version (b) with using \(\beta_{0,s_\text{posterior}}\),
\(\beta_{1,s_\text{posterior}}\) and
\(\sigma_{\gamma,s_\text{posterior}}\),

(3\textsuperscript{rd} column)
\(\textsf{E}_{(c,A,G),s_\text{posterior}}\): Sample mean of posterior
prediction on reference grid structure with Gaussian sampling for
\(\gamma_i\),

(4\textsuperscript{th} column)
\(\textsf{E}_{(c,A,u),s_\text{posterior}}\): Sample mean of posterior
prediction on reference grid structure with uncertainty sampling for
\(\gamma_i\),

(5\textsuperscript{th} column)
\(\textsf{E}_{(c,B,G),s_\text{posterior}}\): Sample mean of posterior
prediction on replicate simulation-data structure with gaussian sampling
for \(\gamma_i\),

(6\textsuperscript{th} column)
\(\textsf{E}_{(c,B,u),s_\text{posterior}}\): Sample mean of posterior
prediction on replicate simulation-data structure with uncertainty
sampling for \(\gamma_i\).

\vspace{0mm}

\begin{description}
\item[Simulation Study]
We repeatedly simulate independently from the above build-up \(R=100\)
times. For each simulation run, in total \(N=500\) observations units
and \(G=20\) groups are considered.
\end{description}

Figure \ref{fig-log-link-multilevel-model--A} on page
\pageref{fig-log-link-multilevel-model--A} shows a scatter plot for each
of the simulated groups in the \(r=1\)\textsuperscript{st} simulation
run, with additionally also showing \(E_{(a),s_\text{prior}}\)
\(E_{(b),s_\text{prior}}\) as lines for varying
\(x\in\left[0,2\right]\).

Figure \ref{fig-log-link-multilevel-model--B} in Supplement B shows the
simulated absolute frequencies for all \(20\) groups, \(g=1,\ldots,20\),
for the first \(30\) simulation runs, \(r=1,\ldots, 30\). Figure
\ref{fig-log-link-multilevel-model--SBC} in Supplement B shows the
SBC-results of the simulation study for \emph{Case Study I} by
uniformity check visualizations according to \citet{SaeilynojaEtAl2022}.

\vspace{0mm}

\begin{description}
\item[Results]
Figure \ref{fig-log-link-multilevel-model--C} on page
\pageref{fig-log-link-multilevel-model--C} shows the results of the
simulation study for \emph{Case Study I} by uniformity check
visualizations according to \citet{SaeilynojaEtAl2022}.
\end{description}

We see that \(E_{(a),s_\text{prior}}\) -- results given in the
\emph{1}\textsuperscript{st} row of sub-figures -- only passes the
uniformity check with \(E_{(a),s_\text{posterior}}\). This means that
the conditional perspective is not calibrated with any of the marginal
perspectives analysed here.

\(E_{(b),s_\text{prior}}\) -- \emph{2}\textsuperscript{nd} row -- only
passes the check for \(E_{(b),s_\text{posterior}}\), and
\(E_{(c,A,G),s_\text{posterior}}\). However, from the posterior
application perspective, \(E_{(b),s_\text{posterior}}\) is calibrated
according to \(E_{(b),s_\text{prior}}\), \(E_{(c,A,G),s_\text{prior}}\),
and \(E_{(c,A,u),s_\text{prior}}\).

\(\textsf{E}_{(c,A,G),s_\text{prior}}\) -- \emph{3}\textsuperscript{rd}
row -- passes the check for \(E_{(b),s_\text{posterior}}\),
\(E_{(c,A,G),s_\text{posterior}}\), and
\(E_{(c,B,u),s_\text{posterior}}\).

\(\textsf{E}_{(c,A,u),s_\text{prior}}\) -- \emph{4}\textsuperscript{th}
row -- passes the check for \(E_{(b),s_\text{posterior}}\),
\(E_{(c,A,G),s_\text{posterior}}\), and
\(E_{(c,A,u),s_\text{posterior}}\).

\(\textsf{E}_{(c,B,G),s_\text{prior}}\) -- \emph{5}\textsuperscript{th}
row -- passes the check for \(\textsf{E}_{(c,B,G),s_\text{posterior}}\),
and \(E_{(c,B,u),s_\text{posterior}}\) (with slightly crossing the
boundaries of the theoretical cumulative distribution function (CDF)
region).

\(\textsf{E}_{(c,B,u),s_\text{prior}}\) -- \emph{6}\textsuperscript{th}
row -- passes the check for \(\textsf{E}_{(c,B,G),s_\text{posterior}}\),
and \(E_{(c,B,u),s_\text{posterior}}\).

\subsection*{Case Study II: ANOVA decomposition of a bivariate smooth
effect
surface}\label{case-study-ii-anova-decomposition-of-a-bivariate-smooth-effect-surface}
\addcontentsline{toc}{subsection}{Case Study II: ANOVA decomposition of
a bivariate smooth effect surface}

The 2\textsuperscript{nd} case study uses a QOI-Check of the form
\emph{prior-predictive posterior-derived} in order to the check
\emph{univariate `main' effect functions derived of a bivariate smooth
effect function} as the \emph{quantity of interest} of a generalized
additive models (GAM). The role of the QOI-Check has a stronger focus on
providing support for software implementation checking of a QOI, but
still also has a role in reference population definition and
interpretation checking, as can be seen by the following toy example:

In an applied study, we may consider \emph{mixed species composition }
as a treatment to increase wood-volume of a forest stand. Such a
treatment could be on average effective, meaning that it increases wood
volume at around \(10\) m\textsuperscript{3}/ha for the whole course of
the treatment. At the same time, it could be true that it increases in
broadleaf-conifer mixtures by \(12\) m\textsuperscript{3}/ha and for
broadleaf-broadleaf mixtures by \(8\) m\textsuperscript{3}/ha. Even
though this interaction would be of largest applied interest, it would
still also be correct to conclude that there is a volume increase, on
average, by any treatment. But this `average' is not set in stone: if a
managed stand is to be composed of 3/4 broadleaf-broadleaf mixtures, and
only of 1/4 broadleaf-conifer mixtures, the volume increase is to be
expected at \(9\) m\textsuperscript{3}/ha. That is by using an averaging
operator in order to marginalize an interaction, an inherent population
definition is introduced.

\begin{quote}
``Sometimes it is interesting to specify smooth models with a main
effects \(+\) interaction structure such as
\(\textsf{E}\left(y_i\right) = f_1\left(x_i\right) + f_2\left(z_i\right) + f_3\left(x_i, z_i\right)\)
{[}\ldots{]} for example. Such models should be set up using \texttt{ti}
terms in the model formula. For example:
\texttt{y\ \textasciitilde{}\ ti(x)\ +\ ti(z)\ +\ ti(x,z)} {[}\ldots{]}.
The \texttt{ti} terms produce interactions with the component main
effects excluded appropriately. (There is in fact no need to use
\texttt{ti} terms for the main effects here, \texttt{s} terms could also
be used.)'' \hfill \citep[p.~79]{Wood2024}
\end{quote}

For an \emph{additive model} with two continuous covariates
\(x\in\left[0,1\right]\) and \(z\in\left[0,1\right]\), we can write a
linear predictor for a bivariate `effect surface'
\(f\left(x_i,z_i\right)\) in an ANOVA decomposed form as
\citep[p.~8]{Gu2002}: \[
\eta_i=f\left(x_i,z_i\right)=f_\varnothing+f_{x}\left(x_i\right)+f_{z}\left(z_i\right)+f_{x,z}\left(x_i,z_i\right)
\] with: \begin{eqnarray*}
f_\varnothing&=&\int_0^1\int_0^1\left(f\left(x,z\right)p\left(x\right)\partial x\right)p\left(z\right)\partial z\\
f_x\left(x_i\right)&=&\int_0^1p\left(z\right)f\left(x_i,z\right)\partial z-f_\varnothing\\
f_z\left(z_i\right)&=&\int_0^1p\left(x\right)f\left(x,z_i\right)\partial x-f_\varnothing\\
f_{x,z}\left(x_i,z_i\right)&=&f\left(x_i,z_i\right)-f_x\left(x_i\right)-f_z\left(z_i\right)+f_\varnothing\\
\end{eqnarray*} That is, we are using a decomposition into a global
intercept \(f_\varnothing\), two univariate centered main effect
functions, \(f_x\left(x\right)\), and \(f_z\left(z\right)\), and a
bivariate interaction surface, \(f_{x,z}\left(x,z\right)\). This split
is achieved by using integration as \emph{averaging operators}
\citep{Gu2002}, with including the \emph{marginal probability density
within a reference population definition}, \(p\left(x\right)\),
\(p\left(z\right)\), into the integral.

So by this case study, we aim to decompose a bivariate smooth effect
into individual ANOVA-like components, which is not directly implemented
at the modeling stage by the \emph{brms} package. Therefore, we must
implement this decomposition post estimation, which introduces the need
for implementation checking (using the QOI-Check). It is realized quite
quickly that the necessary steps won't be implemented just within a few
lines of programming code, and applied researchers therefore need to
withstand Murphy's law.

At the interpretation stage, we need to make sure to know the answer on:
\emph{Which population definition and averaging operator underlies the
main effect functions \(f_1\left(x\right)\) and \(f_2\left(z\right)\)?}

To me, there appears to be no right or wrong here, but it is clearly
essential that researchers know what their interpretation is based on!

In the R add-on package \emph{brms} \citep{Buerkner2017, Buerkner2018},
smooth terms, such as the \deleted{following} call \texttt{s(x,\ z)} for
estimation of a bivariate smooth effect `surface',
\(f\left(x,z\right)\), are implemented using a basis function
transformation on the continuous covariates \(x\) and \(z\). The
\emph{brms} default here is a \emph{Thin Plate Regression Spline} basis
\citep{Wood2003} as implemented by the R add-on package \emph{mgcv}
\citep{Wood2011}.

To facilitate posterior sampling, \emph{brms} implements a hierarchical
parameterization for \texttt{s(x,\ z)}, leading to design matrix
\(\mathbf{X}_{f\left(x,z\right)}\) undergoing a sequence of
transformations (see Supplement C.1). For post-estimation calculations,
the design matrix for the prediction `reference grid' data needs to
undergo the exact same transformations, even if the covariate structure
of \(x\) and \(z\) is balanced (very) differently.

So a design matrix for such smooth terms, denoted
\(\mathbf{X}_{f\left(x,z\right)}\), needs to be generated on the
original data \(\mathcal{D}\) in order to conduct posterior sampling.
But for post-estimation calculations on reference data
\(\mathcal{D}_{a},\mathcal{D}_{b},\ldots\), it is then also needed to
generate design-matrices that not only depend on reference data, but
still contain the essential information on the distribution of \(x\) and
\(z\) in the original data \(\mathcal{D}\) such that the posterior
samples can be applied correctly. This can be thought of as if spline
basis would depend on the classical `smoothing spline knot positions',
than the post processing on reference data needs to incorporate the knot
positions as originally generated for \(\mathcal{D}\) in order to
correctly link spline basis function with estimated parameters.

This leads to a chain of implemented software units that needs to be
checked for correctness during post-estimation calculation of QOIs. The
details of this implementation are described in Supplement B.1.

\vspace{0mm}

\begin{description}
\item[Covariate Structure \(\mathcal{D}\) (as used during Posterior
Sampling)]
The covariates \(x\) and \(z\) that are plugged into \texttt{s(x,z)} are
generated by two independent Beta distributions: \[
\mu_x=\frac{\exp\left(\beta_{0,x}\right)}{1+\exp\left(\beta_{0,x}\right)},
x\sim\text{Beta}\left(\mu_x,\phi_{x}\right),
\text{ and }
\mu_z=\frac{\exp\left(\beta_{0,z}\right)}{1+\exp\left(\beta_{0,z}\right)},
z\sim\text{Beta}\left(\mu_z,\phi_{z}\right).
\]
\item[\vspace{0mm}Likelihood Model]
We use a Normal response model with identity link-function: \[
Y_{i}\sim\text{Normal}\left(\mu_{i},\sigma^2\right),\,\,\mu_{i}=\eta_{i},\,\,\eta_{i}=\beta_0+f\left(x_i,z_i\right).
\] Within one single estimation framework, we also estimate the
distribution of \(x\) and \(z\) based on the above data-generating
model, i.e.~we additionally estimate
\(\left(\beta_{0,x},\phi_{x},\beta_{0,z},\phi_{z}\right)^\top\):
\end{description}

\begin{Shaded}
\begin{Highlighting}[]
\NormalTok{frmla }\OtherTok{\textless{}{-}} \FunctionTok{bf}\NormalTok{(x }\SpecialCharTok{\textasciitilde{}} \DecValTok{1}\NormalTok{, }\AttributeTok{family =} \FunctionTok{Beta}\NormalTok{(}\AttributeTok{link =} \StringTok{"logit"}\NormalTok{)) }\SpecialCharTok{+}
  \FunctionTok{bf}\NormalTok{(z }\SpecialCharTok{\textasciitilde{}} \DecValTok{1}\NormalTok{, }\AttributeTok{family =} \FunctionTok{Beta}\NormalTok{(}\AttributeTok{link =} \StringTok{"logit"}\NormalTok{)) }\SpecialCharTok{+}
  \FunctionTok{bf}\NormalTok{(y }\SpecialCharTok{\textasciitilde{}} \FunctionTok{s}\NormalTok{(x, z, }\AttributeTok{k =} \DecValTok{10}\NormalTok{), }\AttributeTok{family =} \FunctionTok{gaussian}\NormalTok{())}
\end{Highlighting}
\end{Shaded}

\vspace{0mm}

\begin{description}
\item[Prior Specification]
We set the following prior distribution specifications:
\(\beta_{0,x} \sim \text{Normal}\left(0, 0.001^2\right)\),
\(\beta_{0,z} \sim \text{Normal}\left(-1.1, 0.001^2\right)\),
\(\beta_{0,y} \sim \text{Normal}\left(0.5, 0.1^2\right)\),
\(\beta_{\texttt{s(x,z)},1} \sim \text{Normal}\left(1, 0.1^2\right)\),
\(\beta_{\texttt{s(x,z)},2} \sim \text{Normal}\left(-1, 0.1^2\right)\),
\(\sigma_{\texttt{s(x,z)}} \sim \text{Normal}_{+}\left(1, 0.1^2\right)\),
\(\phi_{x} \sim \text{Normal}_{+}\left(3, 0.001^2\right)\),
\(\phi_{z} \sim \text{Normal}_{+}\left(3, 0.001^2\right)\),
\(\sigma_{y} \sim \text{Normal}_{+}\left(0.1, 0.01^2\right)\), where
\(\text{Normal}_{+}\) denotes the half-normal distribution truncated to
only positive real numbers.
\item[\vspace{0mm}Quantity of Interest \emph{ANOVA Decomposition of
Bivariate Spline}]
For a bivariate smooth effect such as used in generalized additive
models as provided by the famous R add-on package \emph{mgcv}
\citep{Wood2011}, we aim to specify a `main effects + interaction'
during post precessing of posterior samples: \[
\text{E}\left(y_i\right)=f_\varnothing+f_{x}\left(x_i\right)+f_{z}\left(z_i\right)+f_{x,z}\left(x_i,z_i\right),
\] with the roles of these single components as described above. The
interpretation of the resulting main effect functions depends on the
averaging operatior / `weighting scheme':
\end{description}

\begin{enumerate}
\def\labelenumi{(\alph{enumi})}
\item
  \begin{description}
  \item[with]
  consideration of the empirical distribution of the respective other
  variable:
  \[f_v\left(v_i\right)=\int_0^1p\left(v^\prime\right)f\left(v_i,v^\prime\right)\partial v^\prime-f_\varnothing,\,v\in\left\{x,z\right\},v\neq v^\prime,\]
  or
  \end{description}
\item
  \begin{description}
  \item[without]
  consideration of the empirical distribution of the respective other
  variable:
  \[f_v\left(v_i\right)=\int_0^1f\left(v_i,v^\prime\right)\partial v^\prime-f_\varnothing,\,v\in\left\{x,z\right\},v\neq v^\prime.\]
  \end{description}
\end{enumerate}

The QOI-Check helps to clarify which weighting scheme is being used and
whether that is in accordance with the researchers intentions.

\vspace{0mm}

\begin{description}
\item[Prior / Posterior Data-Structure]
We set up a data-set in order to for generation of prior or posterior
predictions by establishing an equidistant grid on \((x,z)\). For
version (b), we don't use a distribution weight -- ie., we apply a
uniform weight for each value pair \(\left(x_i, z_i\right)\), and for
version (a) we use the estimated univariate distribution as weighting
function \(p\left(v\right)\), \(v\in\left\{x,z\right\}\).
\end{description}

In the code -- see Supplement C -- I call this \emph{reference grid
prior prediction}. Here, \(N = 10000\) points \(\left(x_i, z_i\right)\)
are simulated exactly as in the initial data posterior estimation data
\(\mathcal{D}\). However, the values for \(x\) are overwritten --
sequentially -- by fixed values, e.g.~\(x = 0.25\). Further, parameters
are simulated from the prior distribution. Then the design matrix is
subjected to the transformations described above, and by this, in
combination with the prior samples, we arrive at prior predictions. The
sample mean from this is the prior predictive QOI for the respective
fixed value of \(x\). These values\hspace{0pt}are then compared with two
versions of the posterior QOI, once with the unweighted calculation of
the conditional expectation -- on an equidistant grid of \(x\) and \(z\)
--, and once weighted according to the distribution of \(z\), which is
also estimated using model parameters, as described above.

\vspace{0mm}

\begin{description}
\item[Simulation Study]
We repeatedly simulate independently from the above build-up \(R=20\)
times.
\item[\vspace{0mm}Results]
Figure \ref{fig-beta-beta-gaussian-additive-model--A} on page
\pageref{fig-beta-beta-gaussian-additive-model--A} shows the uniformity
check for the first main effect \(f\left(x\right)\) at a sequence of
values for \(x\) for version (a) with consideration of the empirical
distribution of \(z\). Values that don't pass the check seem practically
negligible, also considering the multiple testing attribute of this
check.
\end{description}

\section{Conclusion}\label{conclusion}

A systematic check for quantities of interest as derived of a Bayesian
probabilistic model was introduced. This QOI-Check is supposed to be an
addition to the Bayesian workflow, specifically addressing the need to
validate post-estimation calculations of derived quantities and their
interpretation. This method is particularly important when dealing with
complex models and QOIs that are not directly represented directly by
single model parameters.

While existing Bayesian model checking tools, such as PPCs, HPCs, and
SBC, are essential for ensuring model validity, they do not directly
address the challenges of validating post-estimation derived quantities
of interest (QOIs). \citet{ModrakEtAl2023} already introduced SBC with
derived quantities, but without a focus on predictive quantities, and
therefore not on new reference grid data structures. The QOI-Check is
developed to fill this gap by focusing specifically on the
implementation and interpretation of QOIs.

It was demonstrated on two application examples (`Case Studies') that
this check not only leads to trustful post-estimation computations, but
also improves how the respective QOI aligns with a specific population
definition. Case Study I focused on understanding the underlying
population definition for a marginal perspective on the expected value
in a rather simple regression model using a non-linear link function.
Case Study II demonstrates the application of the QOI-Check to a more
complex software implementation task involving ANOVA decomposition of a
bivariate smooth effect surface, therefore having a stronger focus on
the computational implementation of the QOI.

\section*{References}\label{references}
\addcontentsline{toc}{section}{References}

\renewcommand{\bibsection}{}
\bibliography{references.bib}

\clearpage

\section*{Figures}\label{figures}
\addcontentsline{toc}{section}{Figures}

\begin{figure}[!h]
\centering
\begin{subfigure}{0.99\textwidth}
\centering
\begin{tikzpicture}[node distance=1cm, rounded corners]
\tikzstyle{box} = [draw, minimum width=10.6cm, minimum height=4.2cm]
\node[box, fill = lightgray] at (0, -3.1) {};
\tikzstyle{box} = [draw, minimum width=10.9cm, minimum height=4.9cm]
\node[box] at (0, -2.9) {};
\tikzstyle{box} = [draw, minimum width=10.3cm, minimum height=1.5cm]
\node[box, dashed] at (0, -4.3) {};
\node[rectangle, draw, text centered, minimum height=0cm, minimum width=3cm] at (0, 0.5) 
  {Sample \( \boldsymbol{\theta}_1, \dots, \boldsymbol{\theta}_R \) from \( p(\boldsymbol{\theta}) \), with $\boldsymbol{\theta}=\left(\theta_1,\ldots,\theta_p\right)^\top$.};
\node at (0, -.7) {\textcolor{gray}{\textbf{\textsf{Repeat}}} for each $r\in\left\{1,\ldots,R\right\}$:};
\node[rectangle, draw, fill = white, text centered, minimum height=0cm, minimum width=3cm] at (0, -1.4) 
  {Simulate \( \mathbf{y}_r \) given \( \boldsymbol{\theta}_r \).};
\node[rectangle, draw, fill = white, text centered, minimum height=0cm, minimum width=3cm] at (0, -2.6) 
  {Sample \( \boldsymbol{\theta}_{r,1}, \dots, \boldsymbol{\theta}_{r,S} \) from \( p(\boldsymbol{\theta} \mid \mathbf{y}_r) \).};
\node at (0, -3.9) {For each $j\in\left\{1,\ldots,p\right\}$:};
\node[rectangle, draw, fill = white, text centered, minimum height=0cm, minimum width=3cm] at (0, -4.55) 
  {Return cardinal number of $\left\{s\in\left\{1,\ldots,S\right\}:\,\,\theta_{j,r}<\theta_{j,r,s}\right\}$.};
\draw[->] (0,.05) -- (0,-.35);
\draw[->] (0,-1.8) -- (0,-2.2);
\draw[->] (0,-3.0) -- (0,-3.4);
\draw[->] (0,-5.45) -- (0,-5.85);
\node[rectangle, draw, text centered, minimum height=0cm, minimum width=3cm] at (0, -6.25) 
  {For each $j\in\left\{1,\ldots,p\right\}$: Uniformity check of cardinal numbers.};
\end{tikzpicture}
\caption{SBC following \cite{TaltsEtAl2018}, comparing directly a parameter prior sample to the respective parameter posterior samples of that simulation run $r$. The 'cardinal number' of a set is the number of its elements, e.g. the cardinal number of set $A=\left\{a_1, a_2, a_3\right\}$ is $3$, denoted $|A|=3$.}
\end{subfigure}
\hfill
\begin{subfigure}{0.99\textwidth}
\centering
\begin{tikzpicture}[node distance=1cm, rounded corners]
\tikzstyle{box} = [draw, minimum width=1, minimum height=1cm]
\node[] at (0, 1.5) {}; 
\tikzstyle{box} = [draw, minimum width=12.2cm, minimum height=4.2cm]
\node[box, fill = lightgray] at (0, -3.1) {};
\tikzstyle{box} = [draw, minimum width=12.5cm, minimum height=4.9cm]
\node[box] at (0, -2.9) {};
\tikzstyle{box} = [draw, minimum width=11.9cm, minimum height=1.5cm]
\node[box, dashed] at (0, -4.3) {};
\node[rectangle, draw, text centered, minimum height=0cm, minimum width=3cm] at (0, 0.5) 
  {Sample \( \boldsymbol{\theta}_1, \dots, \boldsymbol{\theta}_R \) from \( p(\boldsymbol{\theta}) \), with $\boldsymbol{\theta}=\left(\theta_1,\ldots,\theta_p\right)^\top$.};
\node at (0, -.7) {\textcolor{gray}{\textbf{\textsf{Repeat}}} for each $r\in\left\{1,\ldots,R\right\}$:};
\node[rectangle, draw, fill = white, text centered, minimum height=0cm, minimum width=3cm] at (0, -1.4) 
  {Simulate \( \mathbf{y}_r \) given \( \boldsymbol{\theta}_r \).};
\node[rectangle, draw, fill = white, text centered, minimum height=0cm, minimum width=3cm] at (0, -2.6) 
  {Sample \( \boldsymbol{\theta}_{r,1}, \dots, \boldsymbol{\theta}_{r,S} \) from \( p(\boldsymbol{\theta} \mid \mathbf{y}_r) \).};
\node at (0, -3.9) {For each QOI $f_j\left(\boldsymbol{\vartheta}_j\right)$, $\boldsymbol{\vartheta}_j\subseteq\left\{\theta_k: k=1,\ldots,p\right\}$, $j=1,\ldots,J$:};
\node[rectangle, draw, fill = white, text centered, minimum height=0cm, minimum width=3cm] at (0, -4.55) 
  {Return cardinal number of $\left\{s\in\left\{1,\ldots,S\right\}:\,\,f_j\left(\boldsymbol{\vartheta}_{j,r}\right)<f_j\left(\boldsymbol{\vartheta}_{j,r,s}\right)\right\}$.};
\draw[->] (0,.05) -- (0,-.35);
\draw[->] (0,-1.8) -- (0,-2.2);
\draw[->] (0,-3.0) -- (0,-3.4);
\draw[->] (0,-5.45) -- (0,-5.85);
\node[rectangle, draw, text centered, minimum height=0cm, minimum width=3cm] at (0, -6.25) 
  {For each $j\in\left\{1,\ldots,J\right\}$: Uniformity check of cardinal numbers.};
\end{tikzpicture}
\caption{SBC following \cite{ModrakEtAl2023}, comparing a quantity of interest (QOI) based on the prior with the posterior based analogue, where both versions might also depend on $\mathbf{y}_r$. As already described in slightly more detail in the caption of Subfigure~(a), the 'cardinal number' of a set is the number of its elements.}
\end{subfigure}
\caption{Illustration of \emph{Simulation-Based Calibration (SBC)} check algorithms described in \cite{TaltsEtAl2018} -- shown in Subfigure~(a) --, and \cite{ModrakEtAl2023} -- shown in Subfigure~(b).}
\label{fig-simulation_based_calibration}
\end{figure}

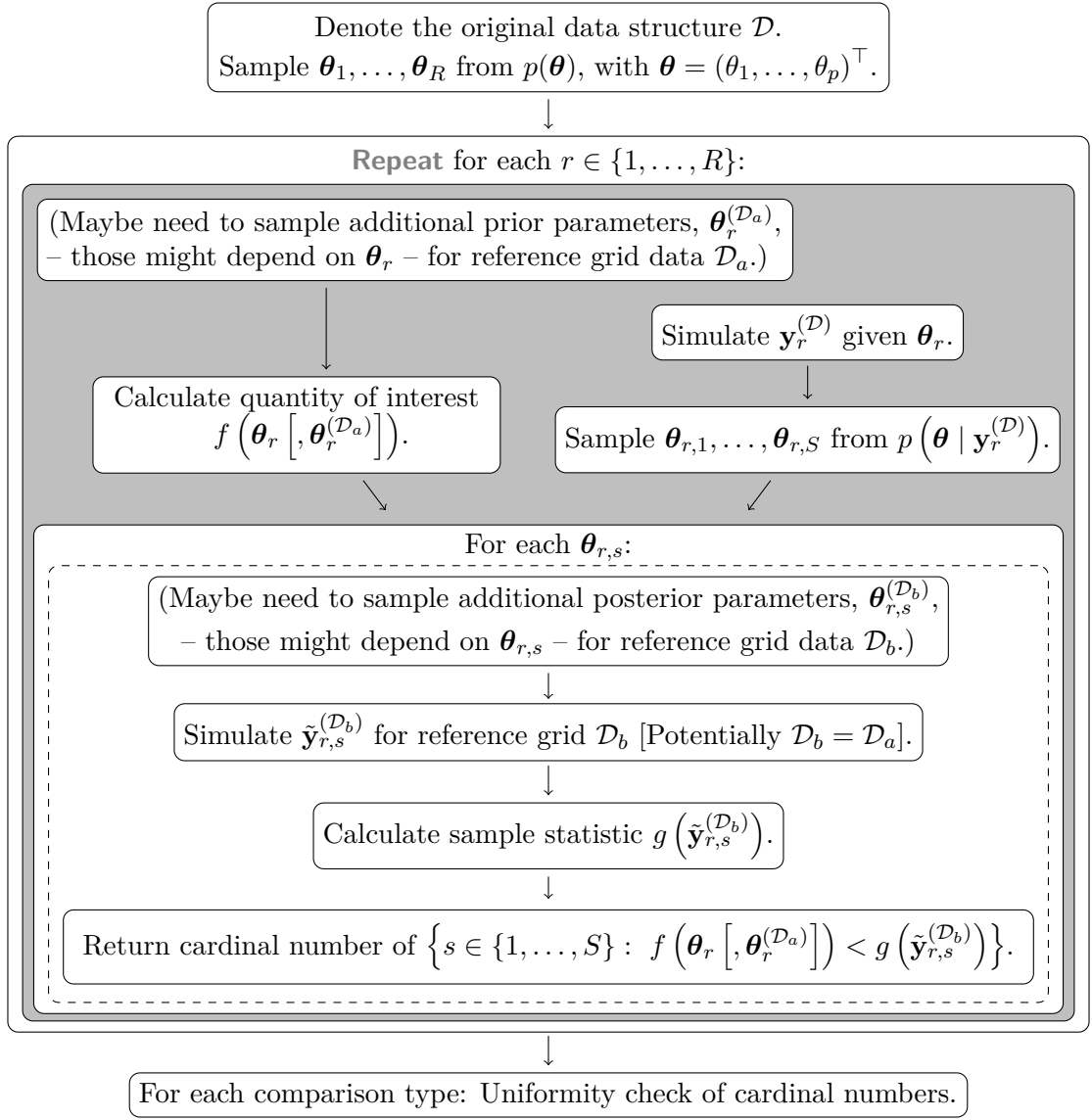
\begin{figure}[!h]
\centering
\begin{tikzpicture}[node distance=1cm, rounded corners]
    \tikzstyle{box} = [draw, minimum width=14.6cm, minimum height=12.1cm]
    \node[box] at (0, -6.5) {};
    \tikzstyle{box} = [draw, minimum width=14.2cm, minimum height=11.3cm]
    \node[box, fill = lightgray] at (0, -6.75) {};
    \tikzstyle{box} = [draw, minimum width=13.9cm, minimum height=6.6cm]
    \node[box, fill = white] at (0, -9) {};
    \tikzstyle{box} = [draw, minimum width=13.5cm, minimum height=5.9cm]
    \node[box, dashed] at (0, -9.2) {};
  
    
    \tikzstyle{box} = [draw, minimum width=9.2cm, minimum height=1.2cm]
    \node[box, fill = white] at (0, .75) {};
    \node[text centered, minimum height=0cm, minimum width=3cm] at (0, 1) 
        {Denote the original data structure $\mathcal{D}$.};
    \node[text centered, minimum height=0cm, minimum width=3cm] at (0, 0.5) 
        {Sample \( \boldsymbol{\theta}_1, \dots, \boldsymbol{\theta}_R \) from \( p(\boldsymbol{\theta}) \), with $\boldsymbol{\theta}=\left(\theta_1,\ldots,\theta_p\right)^\top$.};    
        
    \node at (0, -.8) {\textcolor{gray}{\textbf{\textsf{Repeat}}} for each $r\in\left\{1,\ldots,R\right\}$:};

    \tikzstyle{box} = [draw, minimum width=10.2cm, minimum height=1.1cm]
    \node[box, fill = white] at (-1.8, -1.85) {};
    \node[text centered, minimum height=0cm, minimum width=3cm] at (-1.8, -1.6) 
        {(Maybe need to sample additional prior parameters, $\boldsymbol{\theta}_{r}^{\left(\mathcal{D}_a\right)}$, };
    \node[text centered, minimum height=0cm, minimum width=3cm] at (-1.9, -2.1) 
        {-- those might depend on $\boldsymbol{\theta}_r$ -- for reference grid data $\mathcal{D}_a$.)};
    \tikzstyle{box} = [draw, minimum width=5.5cm, minimum height=1.3cm]
    \node[box, fill = white] at (-3.4, -4.35) {};
    \node[] at (-3.4, -4){Calculate quantity of interest};
    \node[] at (-3.2, -4.5){$f\left(\boldsymbol{\theta}_{r}\left[,\boldsymbol{\theta}_{r}^{\left(\mathcal{D}_a\right)}\right]\right)$.};
    
    \node[rectangle, draw, fill = white, text centered, minimum height=0cm, minimum width=3cm] at (3.5, -3.1) 
        {Simulate $\mathbf{y}_r^{\left(\mathcal{D}\right)}$ given $\boldsymbol{\theta}_r$.};
    \node[rectangle, draw, fill = white, text centered, minimum height=0cm, minimum width=3cm] at (3.5, -4.55) 
        {Sample $\boldsymbol{\theta}_{r,1}, \dots, \boldsymbol{\theta}_{r,S}$ from $ p\left(\boldsymbol{\theta} \mid \mathbf{y}_r^{\left(\mathcal{D}\right)}\right)$.};

    \node[] at (0, -6){For each $\boldsymbol{\theta}_{r,s}$:};
    \tikzstyle{box} = [draw, minimum width=10.8cm, minimum height=1.2cm]
    \node[box, fill = white] at (0, -7.0) {};
    \node[text centered, minimum height=0cm, minimum width=3cm] at (-0, -6.7) 
        {(Maybe need to sample additional posterior parameters, $\boldsymbol{\theta}_{r,s}^{\left(\mathcal{D}_b\right)}$, };
    \node[text centered, minimum height=0cm, minimum width=3cm] at (-0, -7.3) 
        {-- those might depend on $\boldsymbol{\theta}_{r,s}$ -- for reference grid data $\mathcal{D}_b$.)};
    \node[rectangle, draw, text centered, minimum height=0cm, minimum width=3cm] at (0, -8.5){Simulate $\tilde{\mathbf{y}}_{r,s}^{\left(\mathcal{D}_b\right)}$ for reference grid $\mathcal{D}_b$ [Potentially $\mathcal{D}_b=\mathcal{D}_a$].};
    \node[rectangle, draw, text centered, minimum height=0cm, minimum width=3cm] at (0, -9.9){Calculate sample statistic $g\left(\tilde{\mathbf{y}}_{r,s}^{\left(\mathcal{D}_b\right)}\right)$.};

    \tikzstyle{box} = [draw, minimum width=13.1cm, minimum height=1cm]
    \node[box] at (0, -11.4) {};
    \node[text centered, minimum height=0cm, minimum width=3cm] at (0, -11.4) 
        {Return cardinal number of $\left\{s\in\left\{1,\ldots,S\right\}:\,\,
        f\left(\boldsymbol{\theta}_{r}\left[,\boldsymbol{\theta}_{r}^{\left(\mathcal{D}_a\right)}\right]\right)
        <
        g\left(\tilde{\mathbf{y}}_{r,s}^{\left(\mathcal{D}_b\right)}\right)\right\}$.};
    %
    \node[rectangle, draw, text centered, minimum height=0cm, minimum width=3cm] at (0, -13.4) 
        {For each comparison type: Uniformity check of cardinal numbers.};
    \draw[->] (0,.05) -- (0,-.35);
    \draw[->] (-3,-2.5) -- (-3,-3.6);
    \draw[->] (-2.5,-5.1) -- (-2.2,-5.5);
    \draw[->] (3.5,-3.55) -- (3.5,-4);
    \draw[->] (3,-5.1) -- (2.7,-5.5);
    \draw[->] (0,-7.7) -- (0,-8.05);
    \draw[->] (0,-8.95) -- (0,-9.35);
    \draw[->] (0,-10.45) -- (0,-10.75);
    \draw[->] (0,-12.6) -- (0,-13.0);

\end{tikzpicture}
\caption{\emph{Prior-Derived Posterior-Predictive Consistency Check}.}
\label{fig:Prior-Derived-Posterior-Predictive}
\end{figure}

\begin{figure}[!h]
\centering
\begin{tikzpicture}[node distance=1cm, rounded corners]
    \tikzstyle{box} = [draw, minimum width=14.7cm, minimum height=12.1cm]
    \node[box] at (0, -6.5) {};
    \tikzstyle{box} = [draw, minimum width=14.3cm, minimum height=11.3cm]
    \node[box, fill = lightgray] at (0, -6.75) {};
    \tikzstyle{box} = [draw, minimum width=13.9cm, minimum height=5.75cm]
    \node[box, fill = white] at (0, -9.4) {};
    \tikzstyle{box} = [draw, minimum width=13.5cm, minimum height=4.9cm]
    \node[box, dashed] at (0, -9.7) {};

    \tikzstyle{box} = [draw, minimum width=9.2cm, minimum height=1.2cm]
    \node[box, fill = white] at (0, .75) {};
    \node[text centered, minimum height=0cm, minimum width=3cm] at (0, 1) 
        {Denote the original data structure $\mathcal{D}$.};
    \node[text centered, minimum height=0cm, minimum width=3cm] at (0, 0.5) 
        {Sample \( \boldsymbol{\theta}_1, \dots, \boldsymbol{\theta}_R \) from \( p(\boldsymbol{\theta}) \), with $\boldsymbol{\theta}=\left(\theta_1,\ldots,\theta_p\right)^\top$.};
        
    \node at (0, -.8) {\textcolor{gray}{\textbf{\textsf{Repeat}}} for each $r\in\left\{1,\ldots,R\right\}$:};

    \tikzstyle{box} = [draw, minimum width=10.2cm, minimum height=1.2cm]
    \node[box, fill = white] at (-1.8, -1.85) {};
    \node[text centered, minimum height=0cm, minimum width=3cm] at (-1.8, -1.6) 
        {(Maybe need to sample additional prior parameters, $\boldsymbol{\theta}_{r}^{\left(\mathcal{D}_a\right)}$, };
    \node[text centered, minimum height=0cm, minimum width=3cm] at (-1.7, -2.1) 
        {-- those might depend on $\boldsymbol{\theta}_r$ -- for reference grid data $\mathcal{D}_a$.)};
    \draw[->] (0.3,-2.6) -- (0.5,-3.1) -- (-0.1,-3.2);
    \tikzstyle{box} = [draw, minimum width=5.5cm, minimum height=1.2cm]
    \node[box, fill = white] at (-2.95, -3.5) {};
    \node[] at (-2.95, -3.25){Simulate prior-predicted $\tilde{\mathbf{y}}_{r}^{\left(\mathcal{D}_a\right)}$,};
    \node[] at (-2.9, -3.8){for reference grid data $\mathcal{D}_a$.};
    \draw[->] (-0.1,-3.8) -- (0.1,-3.9) -- (-0.4,-4.6);
    \tikzstyle{box} = [draw, minimum width=4.6cm, minimum height=1.2cm]
    \node[box, fill = white] at (-2.8, -5.1) {};
    \node[] at (-2.8, -4.8){Calculate sample statistic};
    \node[] at (-2.6, -5.3){$f\left(\tilde{\mathbf{y}}_{r}^{\left(\mathcal{D}_a\right)}\right)$.};

    \node[rectangle, draw, fill = white, text centered, minimum height=0cm, minimum width=3cm] at (4, -3.8) 
        {Simulate \(\mathbf{y}_r^{\mathcal{D}} \) given \( \boldsymbol{\theta}_r \).};
    \node[rectangle, draw, fill = white, text centered, minimum height=0cm, minimum width=3cm] at (3.6, -5.2) 
        {Sample $\boldsymbol{\theta}_{r,1}, \ldots, \boldsymbol{\theta}_{r,S}$ from $p\left(\boldsymbol{\theta} \mid \mathbf{y}_r^{\mathcal{D}}\right)$.};

    \node[] at (0, -6.9){For each $\boldsymbol{\theta}_{r,s}$:};
    
    
    \tikzstyle{box} = [draw, minimum width=10.8cm, minimum height=1.3cm]
    \node[box, fill = white] at (0, -8.1) {};
    \node[text centered, minimum height=0cm, minimum width=3cm] at (-0, -7.8) 
        {(Maybe need to sample additional posterior parameters, $\boldsymbol{\theta}_{r,s}^{\left(\mathcal{D}_b\right)}$, };
    \node[text centered, minimum height=0cm, minimum width=3cm] at (-0, -8.4) 
        {-- those might depend on $\boldsymbol{\theta}_{r,s}$ -- for reference grid data $\mathcal{D}_b$.)};

    \node[rectangle, draw, text centered, minimum height=0cm, minimum width=3cm] at (0, -9.9){Calculate quantity of interest $g\left(\boldsymbol{\theta}_{r,s}\left[,\boldsymbol{\theta}_{r,s}^{\left(\mathcal{D}_b\right)}\right]\right)$.};

    \tikzstyle{box} = [draw, minimum width=13cm, minimum height=1cm]
    \node[box] at (0, -11.5) {};
    \node[text centered, minimum height=0cm, minimum width=3cm] at (0, -11.5) 
        {Return cardinal number of $\left\{s\in\left\{1,\ldots,S\right\}:\,\,
        f\left(\tilde{\mathbf{y}}_{r}^{\left(\mathcal{D}_a\right)}\right)
        <
        g\left(\boldsymbol{\theta}_{r,s}\left[,\boldsymbol{\theta}_{r,s}^{\left(\mathcal{D}_b\right)}\right]\right)\right\}$.};
    
    \node[rectangle, draw, text centered, minimum height=0cm, minimum width=3cm] at (0, -13.4) 
        {For each comparison type: Uniformity check of cardinal numbers.};
    \draw[->] (0,.05) -- (0,-.35);
    \draw[->] (4,-4.2) -- (4,-4.7);
    \draw[->] (3,-5.8) -- (2.7,-6.4);
    \draw[->] (-1.5,-5.8) -- (-1.2,-6.4);
    \draw[->] (0,-8.8) -- (0,-9.4);
    \draw[->] (0,-10.4) -- (0,-10.9);
    \draw[->] (0,-12.6) -- (0,-13.0);

\end{tikzpicture}
\caption{\emph{Prior-Predictive Posterior-Derived Consistency Check}.}
\label{fig:Prior-Predictive-Posterior-Derived}
\end{figure}
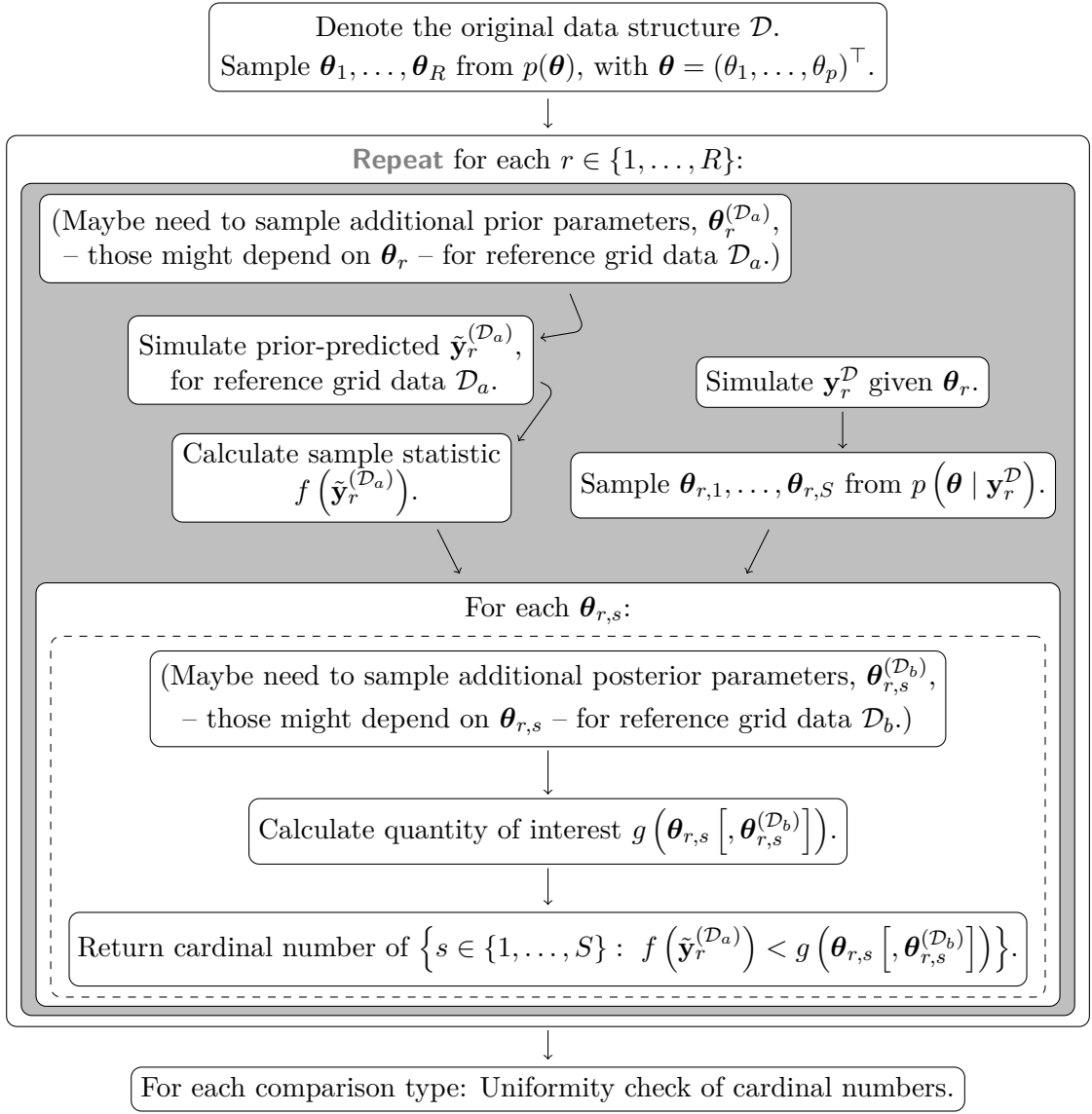

\begin{figure}[!h]
\centering
\begin{subfigure}{0.4\textwidth}
\centering
\begin{tikzpicture}
\tikzstyle{vector} = [align=center]
\tikzstyle{box} = [draw, rounded corners, minimum width=5.6cm, minimum height=3.1cm]
\node[box, fill = lightgray] at (2.2, 0.7) {};
\tikzstyle{box} = [draw, rounded corners, minimum width=0cm, minimum height=0cm]
\node[box, fill = white] at (2.2, 2.55) {Primary simulation data-structure};
\tikzstyle{box} = [draw, sharp corners, minimum width=1.2cm, minimum height=2.5cm]
\node at (0, 2) {\( i,j \)};
\node at (1.4, 2) {\( \mathbf{x} \)};
\node at (2.8, 2) {\( \mathbf{g} \)};
\node at (4.2, 2) {\( \mathbf{y} \)};
\node[] at (0, .5) {\( \begin{array}{c} 1 \\ 2 \\ \vdots \\ N \end{array} \)};
\node[box, fill = white] at (1.4, .5) {\( \begin{array}{c} x_1 \\ x_2 \\ \vdots \\ x_N \end{array} \)};
\node[box, fill = white] at (2.8, .5) {\( \begin{array}{c} g_{\left(1\right)} \\ g_{\left(1\right)} \\ \vdots \\ g_{\left(G\right)} \end{array} \)};
\node[] at (4.2, .5) {\( \begin{array}{c} y_1 \\ y_2 \\ \vdots \\ y_N \end{array} \)};
\end{tikzpicture}
\caption{\emph{Primary data structure}.}
\end{subfigure}
\hfill
\begin{subfigure}{0.5\textwidth}
\centering
\begin{tikzpicture}
\tikzstyle{vector} = [align=center]
\tikzstyle{box} = [draw, rounded corners, minimum width=5.6cm, minimum height=3.1cm]
\node[box, fill = lightgray] at (2.2, -3.3) {};
\tikzstyle{box} = [draw, rounded corners, minimum width=0cm, minimum height=0cm]
\node[box, fill = white] at (2.2, -1.55) {$\mathcal{D}_{(A)}$};
\tikzstyle{box} = [draw, sharp corners, minimum width=1.2cm, minimum height=2.5cm]
\node at (0, -2) {\( i,j \)};
\node at (1.4, -2) {\( \mathbf{x} \)};
\node at (2.8, -2) {\( \mathbf{g} \)};
\node at (4.2, -2) {\( \tilde{\mathbf{y}} \)};
\node[] at (0, -3.5) {\( \begin{array}{c} 1 \\ 2 \\ \vdots \\ N \end{array} \)};
\node[box, fill = white] at (1.4, -3.5) {\( \begin{array}{c} 1 \\ 1 \\ \vdots \\ 1 \end{array} \)};
\node[box, fill = white] at (2.8, -3.5) {\( \begin{array}{c} g_{\left(G+1\right)} \\ g_{\left(G+1\right)} \\ \vdots \\ g_{\left(G+G\right)} \end{array} \)};
\node[] at (4.2, -3.5) {\( \begin{array}{c} \tilde{y}_1 \\ \tilde{y}_2 \\ \vdots \\ \tilde{y}_N \end{array} \)};
\end{tikzpicture}
\caption{\emph{Replicate simulation-data structure} $\mathcal{D}_{(A)}$ where an equally balanced vector -- of length $N$ -- of the \emph{primary simulation data-structure} for $\mathbf{g}$ is re-used for prediction -- vector $\tilde{\mathbf{y}}$ -- of $N$ observation units at $x=1$.}
\end{subfigure}

\begin{subfigure}{0.5\textwidth}
\centering
\begin{tikzpicture}
\tikzstyle{box} = [draw, rounded corners, minimum width=7.4cm, minimum height=3.1cm]
\node[box, fill = lightgray] at (2.9, -3.3) {};
\tikzstyle{box} = [draw, rounded corners, minimum width=0cm, minimum height=0cm]
\node[box, fill = white] at (3, -1.55) {$\mathcal{D}_{(B)}$};
\tikzstyle{box} = [draw, sharp corners, minimum width=1.6cm, minimum height=2.5cm]
\tikzstyle{vector} = [align=center]
\node at (0, -2) {\( i,j \)};
\node at (1.8, -2) {\( \mathbf{x} \)};
\node at (3.6, -2) {\( \mathbf{g} \)};
\node at (5.4, -2) {\( \tilde{\mathbf{y}} \)};
\node[] at (0, -3.5) {\( \begin{array}{c} 1 \\ 2 \\ \vdots \\ N_{(B)} \end{array} \)};
\node[box, fill = white] at (1.8, -3.5) {\( \begin{array}{c} 1 \\ 1 \\ \vdots \\ 1 \end{array} \)};
\node[box, fill = white] at (3.6, -3.5) {\( \begin{array}{c} g_{\left(G+1\right)} \\ g_{\left(G+2\right)} \\ \vdots \\ g_{\left(G+N_{(B)}\right)} \end{array} \)};
\node[] at (5.8, -3.5) {\( \begin{array}{c} \tilde{y}_{(B),1} \\ \tilde{y}_{(B),2} \\ \vdots \\ \tilde{y}_{(B),N_{(B)}} \end{array} \)};
\end{tikzpicture}
\caption{\emph{Reference grid structure} $\mathcal{D}_{(B)}$ where $N_{(B)}$ new observation units are generated, each with $x=1$ and a new grouping variable level that was not included in the $G$ levels present in the posterior estimation data.}
\end{subfigure}
\caption{Schematic illustration of the estimation data and two post-estimation data structures as used and compared in \emph{Case Study I}.}
\label{fig-tikz--A}
\end{figure}
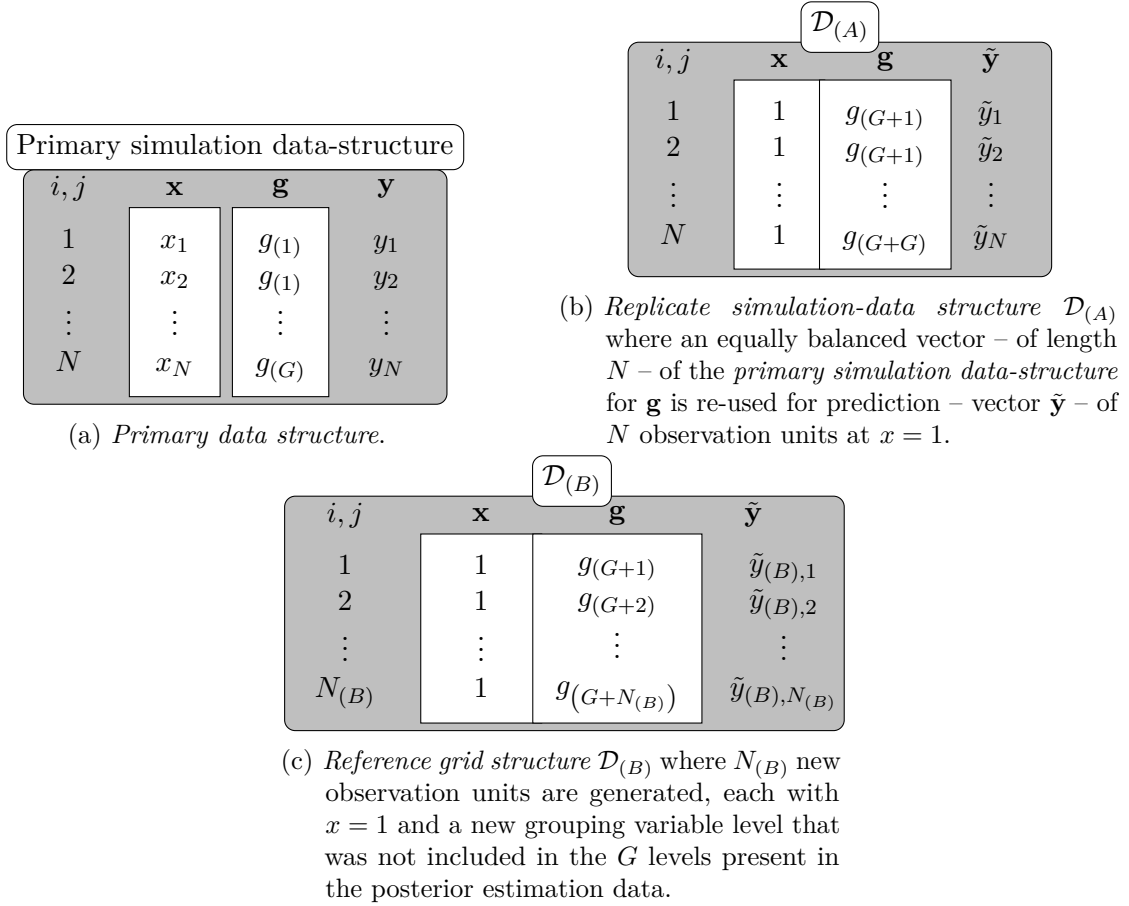

\begin{figure}[!h]
\centering
\includegraphics[width =\textwidth]{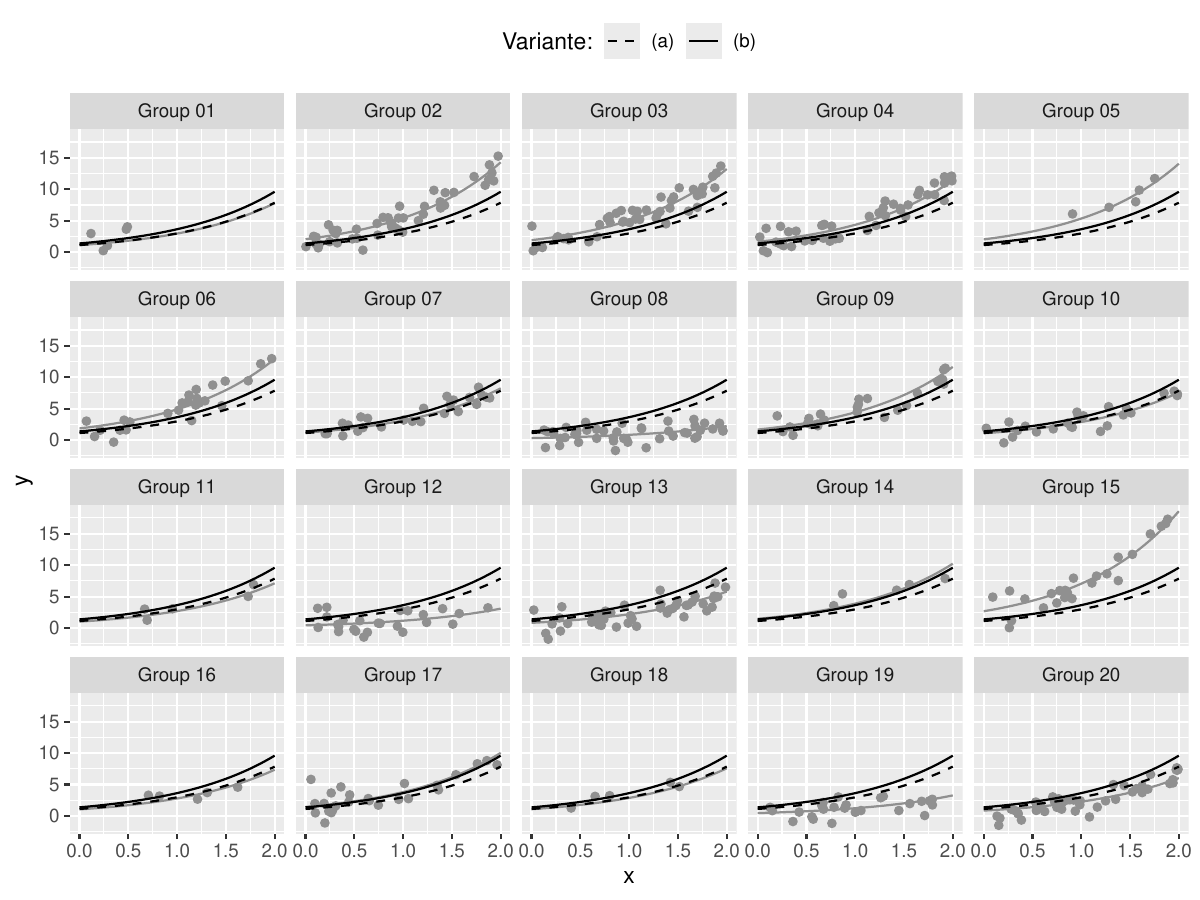}
\caption{Scatterplot of the data in the 1\textsuperscript{st} simulation run (Lines visualize the underlying prior sample $\boldsymbol{\theta}_2=\left(\beta_{0,2},\beta_{1,2},\sigma_{\gamma_{2}},\gamma_{1,2},\ldots,\gamma_{20,2}\right)^\top$). For the \emph{duplicate simulation data-set} the same values of covariates $x$ and $g$ are re-used. }
\label{fig-log-link-multilevel-model--A}
\end{figure}

\begin{figure}[!h]
\centering
\includegraphics[width =\textwidth]{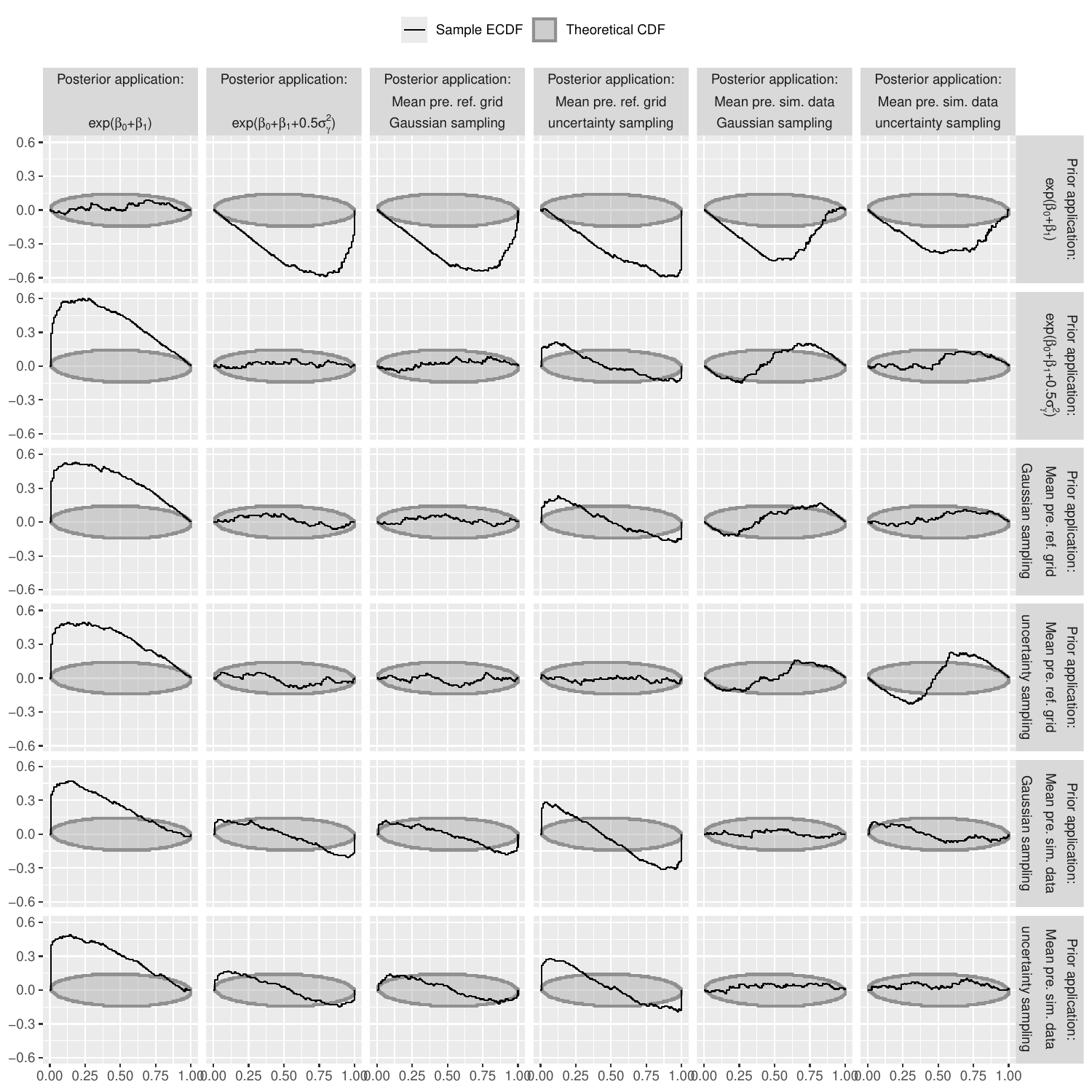}
\caption{Uniformity check visualisations for Prior-Derived Posterior-Predictive and Prior-Predictive Posterior-Derived Consistency Checks for the Quantity of interest in \emph{Case Study I}: Prior-Posterior QOI combinations for which the black line remains inside the theoretical cumulatite distribution function (CDF) pass the check.}
\label{fig-log-link-multilevel-model--C}
\end{figure}

\begin{figure}[!h]
\centering
\includegraphics[width =\textwidth]{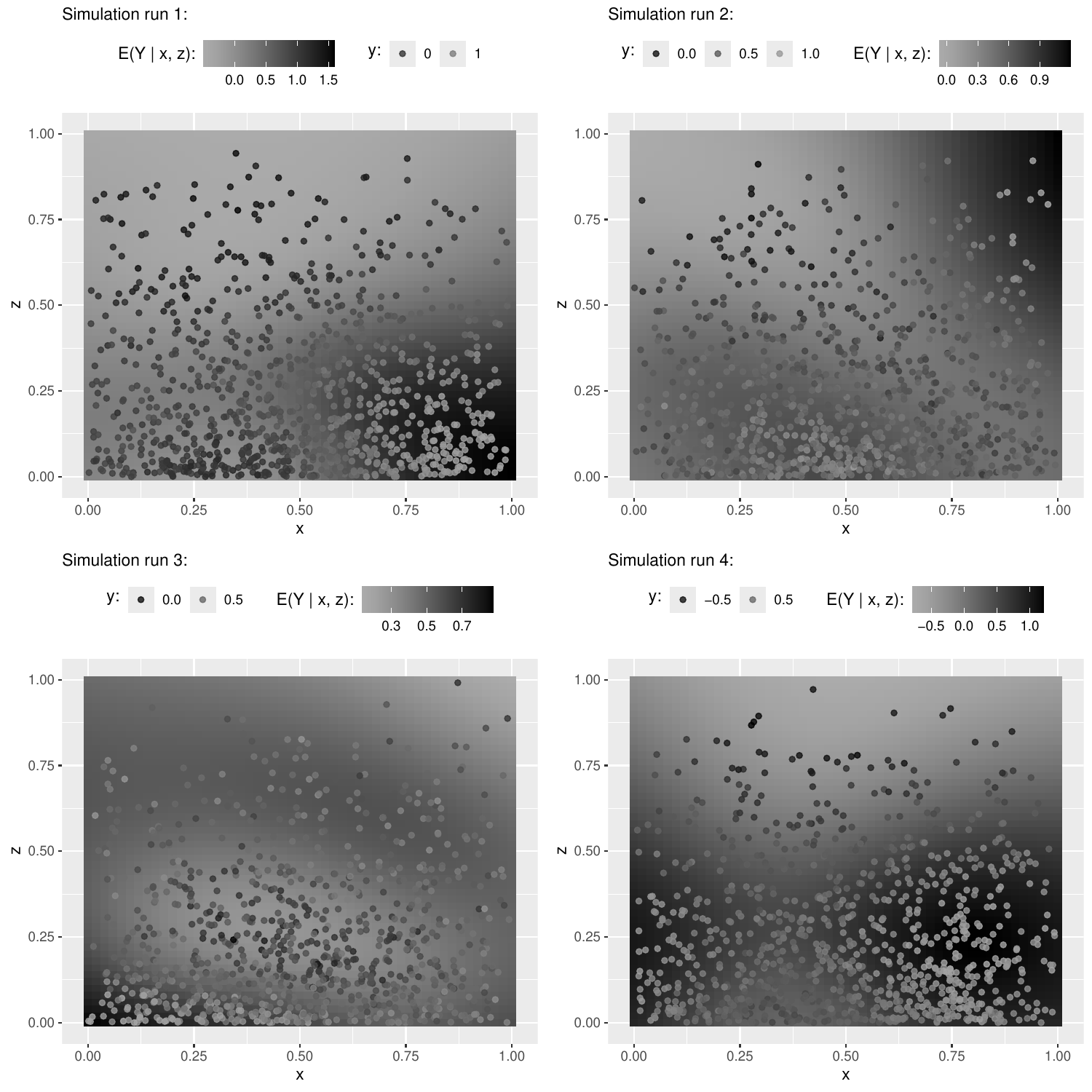}
\caption{\emph{Case Study II}: Data and bivariate smooth estimation of the first four simulation runs, $r\in\left\{1,2,3,4\right\}$.}
\label{fig-case-study-II-bivariate-smooth}
\end{figure}

\begin{figure}[!h]
\centering
\includegraphics[width =\textwidth]{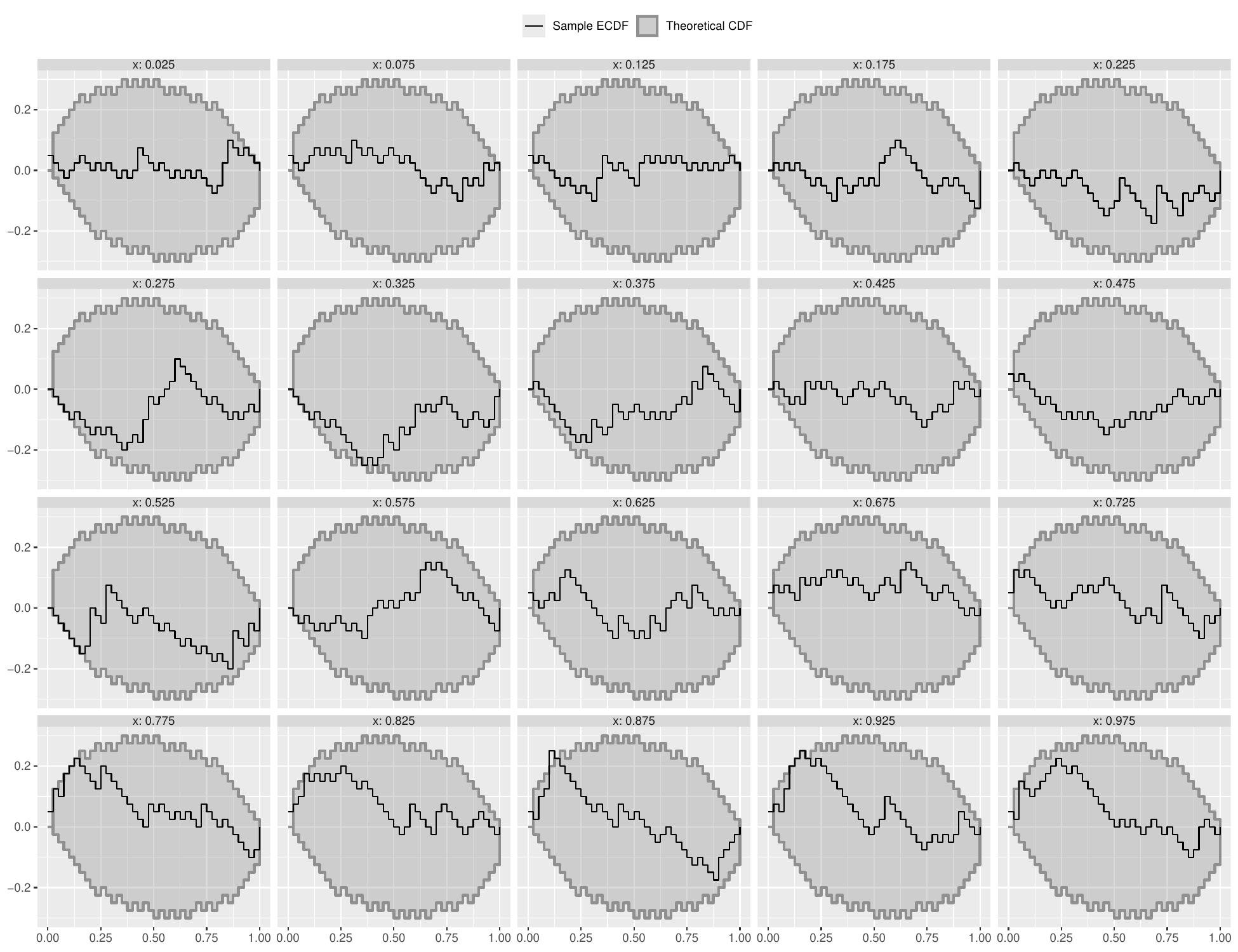}
\caption{Uniformity check visualisations for \emph{Prior-Predictive Posterior-Derived QOI} -- expectation of y conditional on x with marginalizing out the effect of covariate $z$ -- in \emph{Case Study II}: Prior-Posterior QOI combinations for which the black line remains inside the theoretical cumulatite distribution function (CDF) pass the check.}
\label{fig-beta-beta-gaussian-additive-model--A}
\end{figure}

\clearpage

\section*{Supplementary Materials}\label{supplementary-materials}
\addcontentsline{toc}{section}{Supplementary Materials}

\subsection*{Supplement A: Long version of the posterior sampling `toy
algorithm'}\label{supplement-a-long-version-of-the-posterior-sampling-toy-algorithm}
\addcontentsline{toc}{subsection}{Supplement A: Long version of the
posterior sampling `toy algorithm'}

\begin{enumerate}
\def\labelenumi{\arabic{enumi})}
\tightlist
\item
  Sample values of \(\theta_1,\ldots,\theta_S\), from the prior,
  \(p\left(\theta\right)\).
\item
  For each sampled \(\theta_s\), \(s=1,\ldots,S\), a data-set
  \(\tilde{\mathbf{y}}_s\) -- with the same dimension as \(\mathbf{y}\),
  usually conditioned on the same covariate data \(\mathbf{X}\) -- is
  simulated based on the model, \(p\left(Y_i\mid\theta\right)\), or
  \(p\left(Y_i\mid\theta,\mathbf{X}_i\right)\), \(i=1,\ldots,n\). This
  leaves us with \(S\) replicated simulations of the original data-set.
\item
  Define a subset index vector \(\mathcal{S}_{\text{sub}}\), which
  contains a selection of elements of
  \(\mathcal{S}=\left\{1, \ldots,S\right\}\) based on the criterion that
  \(\tilde{\mathbf{y}}_s\), \(s\in\mathcal{S}\), matches exactly --
  after practically meaningful rounding for continuous data -- the
  observed data-set \(\mathbf{y}\). For example, if \(S=5\), and binary
  `matching indicator vector'
  \(\mathbf{I} = \left(1, 0, 1, 0, 1\right)^\top\) -- where each element
  is \(1\) if the generated data \(\tilde{\mathbf{y}}_s\) matches the
  original data, \(\mathbf{y}\), and \(0\) otherwise --, then
  \(\mathcal{S}_{\text{sub}} = \left\{1, 3, 5\right\}\).
\item
  Keep only prior samples \(\theta_s\),
  \(s\in\mathcal{S}_{\text{sub}}\), and, by this, get the distribution
  of prior samples that represents our `knowledge' on the likely
  parameter values that could have generated the observed data vector
  \(\mathbf{y}\).
\end{enumerate}

\citet{Rubin1984} describes this `filtering procedure' as being
`Bayesianly justifiable', since it uses the prior distribution to
generate hypothetical data and then updates this distribution based on
the observed data. This approach aligns with the foundational principles
of Bayesian statistics, allowing for coherent and rational inference
based on prior knowledge and observed data. However, while valid in
theory, it faces the constraint that -- in the usually applied range of
the examined number, \(N\), of observation units, say \(N>50\), and most
often even much larger -- this unstructured search for matching
replicated data-sets is much too inefficient\footnote{Think of the
  simple linear regression model with only one continuous covariate
  \(\mathbf{x}\) where we even can't call out a match as soon as the
  increasingly sorted \(\tilde{\mathbf{y}}_s\) matches the increasingly
  sorted \(\tilde{\mathbf{y}}\) (by application of exchangeability).},
which makes this approach completely impractical when applied in the
real-world.

\begin{figure}[ht]
\centering
\begin{tikzpicture}[node distance=1cm, thick, >=stealth, rounded corners]
    \tikzstyle{box} = [draw, minimum width=7.7cm, minimum height=2cm]
    \node[box, fill = lightgray] at (0, -1.75) {};
    \node[text centered, minimum height=0cm, minimum width=3cm] at (0, 1.5) 
        {We have observed data $\mathbf{y}$.};   
    \node[text centered, minimum height=0cm, minimum width=3cm] at (0, 1) 
        {Initialize \( \mathcal{S} \) as the empty set.};   
    \node[rectangle, draw, text centered, minimum height=0cm, minimum width=3cm] at (0, 3) 
        {Sample \( \boldsymbol{\theta}_1, \dots, \boldsymbol{\theta}_R \) from \( p\left(\boldsymbol{\theta}\right) \).};
    \node (upr) at (0, 1.9) {};
    \node[rectangle, draw, fill = white, text centered, minimum height=0cm, minimum width=3cm] (simulate)  at (0, -1.25) {Simulate \( \tilde{\mathbf{y}}_r \) given \( \boldsymbol{\theta}_r \).};
    \node at (0, -.4) {\textcolor{gray}{\textbf{\textsf{Repeat}}} for each $r\in\left\{1,\ldots,R\right\}$:};
    \node[rectangle, draw, fill = white, text centered, minimum height=0cm, minimum width=3cm, below of=simulate] (subset) 
    {Add $r$ to \( \mathcal{S} \) based on matching \( \tilde{\mathbf{y}}_r \) with \( \mathbf{y} \).};
    
    \node[rectangle, draw, text centered, minimum height=0cm, minimum width=3cm] (keep)  at (0, -4) {Keep \( \boldsymbol{\theta}_s \) for \( s \in \mathcal{S} \)};

    \draw[->] (0,2.6) -- (0,2.1);
    \draw[->] (0,.6) -- (0,0);
    \draw[->] (0, -1.6) -- (0, -1.9);
    \draw[->] (0, -3.1) -- (0, -3.6);
    \tikzstyle{box} = [draw, minimum width=8.1cm, minimum height=5cm]
    \node[box] at (0, -.5) {};
    \tikzstyle{box} = [draw, minimum width=5cm, minimum height=1cm]
    \node[box] at (0, 1.25) {};
\end{tikzpicture}
\caption{\emph{'Naive posterior sampling'} in a simulation study with $R$ replications.}
\label{fig:naive_posterior_sampler}
\end{figure}
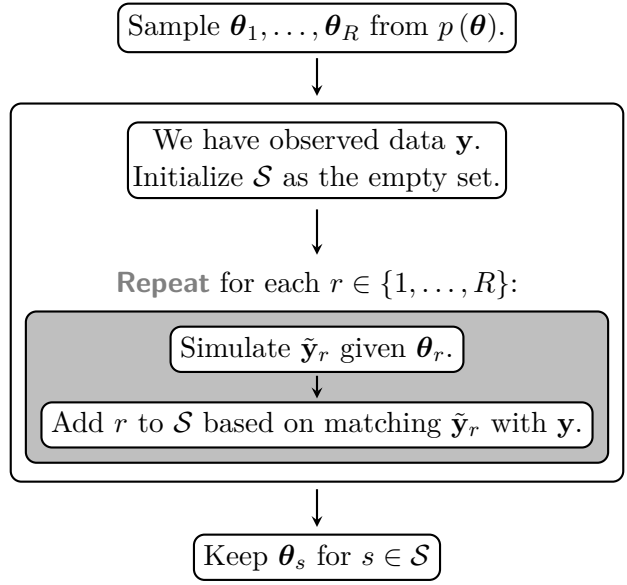

Figure \ref{fig:naive_posterior_sampler} on page
\pageref{fig:naive_posterior_sampler} shows a sketch for naive posterior
sampling -- following \citet[Section 3.1]{Rubin1984} -- in a simulation
study with \(R\) replications.

\subsection*{\texorpdfstring{Supplement B: Supplements to \emph{Case
Study
I}}{Supplement B: Supplements to Case Study I}}\label{supplement-b-supplements-to-case-study-i}
\addcontentsline{toc}{subsection}{Supplement B: Supplements to
\emph{Case Study I}}

\begin{figure}
\centering
\includegraphics[width =\textwidth]{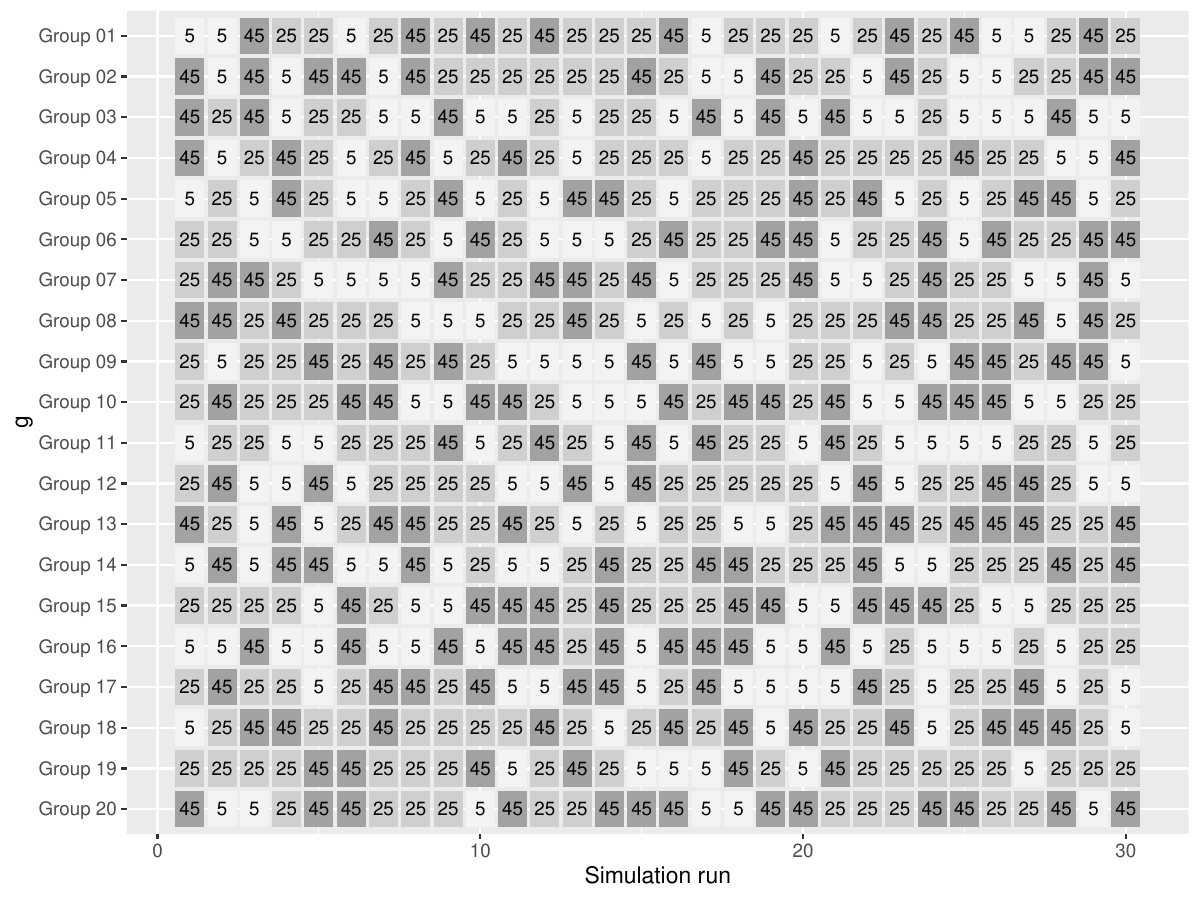}
\caption{Frequencies of simulated levels of grouping variable $g$ in the first 30 simulation runs. }
\label{fig-log-link-multilevel-model--B}
\end{figure}

Figure \ref{fig-log-link-multilevel-model--B} on page
\pageref{fig-log-link-multilevel-model--B} shows the simulated absolute
frequencies for all \(20\) groups, \(g=1,\ldots,20\), for the first
\(30\) simulation runs, \(r=1,\ldots, 30\).

\begin{figure}
\centering
\includegraphics[width =\textwidth]{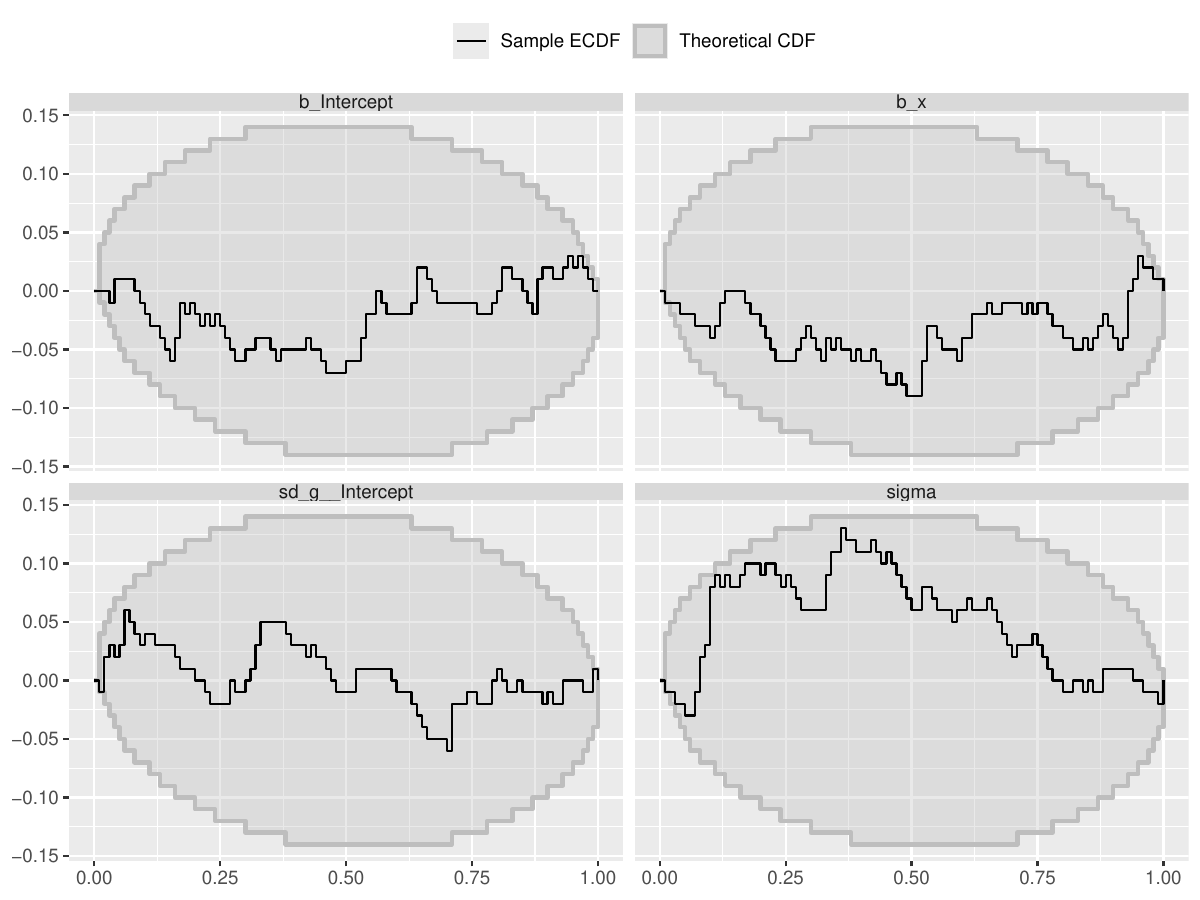}
\caption{SBC for \emph{Case Study I}: All single model parameters pass the check.}
\label{fig-log-link-multilevel-model--SBC}
\end{figure}

Figure \ref{fig-log-link-multilevel-model--SBC} on page
\pageref{fig-log-link-multilevel-model--SBC} shows the SBC-results of
the simulation study for \emph{Case Study I} by uniformity check
visualizations according to \citet{SaeilynojaEtAl2022}.

\subsubsection*{Supplement B.1: R Code}\label{supplement-b.1-r-code}
\addcontentsline{toc}{subsubsection}{Supplement B.1: R Code}

See website
\url{https://uncertaintree.github.io/research/qoi-check.html}.

\subsection*{\texorpdfstring{Supplement C: Supplements to \emph{Case
Study
II}}{Supplement C: Supplements to Case Study II}}\label{supplement-c-supplements-to-case-study-ii}
\addcontentsline{toc}{subsection}{Supplement C: Supplements to
\emph{Case Study II}}

\subsubsection*{\texorpdfstring{Supplement C.1: Smooth Effect
Parametrisation in
\emph{brms}}{Supplement C.1: Smooth Effect Parametrisation in brms}}\label{supplement-c.1-smooth-effect-parametrisation-in-brms}
\addcontentsline{toc}{subsubsection}{Supplement C.1: Smooth Effect
Parametrisation in \emph{brms}}

Matrix \(\mathbf{X}_{\texttt{s(x,z)}}\) for data \(\mathcal{D}\), or
\(\mathcal{D}_a\), \(\ldots\), is constructed using
\texttt{mgcv::PredictMat}. In order to get to random-effects
parameterization, \(\mathbf{X}_{\texttt{s(x,z)}}\) is post-multiplied by
matrix \(\mathbf{U}\), which is a transformation matrix that
orthogonalizes the random effect covariance structure. Then, the columns
of \(\mathbf{X} \mathbf{U}\) are scaled element-wise by the diagonal
entries of the matrix \(\mathbf{D}\), which contains the square roots of
the random effect variances.

The transformed matrix is then split into two components: The component
\(\mathbf{X}_{\texttt{s(x,z)},1}\) corresponds to the columns of
\(\mathbf{X} \mathbf{U}\) that are not penalized (those where the
entries of \(\mathbf{D}\) are zero or not relevant). The `penalized'
components are extracted as \(\mathbf{X}_{\texttt{s(x,z)},2,l}\), where
each of the \(L\) columns, \(\mathbf{x}_{\texttt{s(x,z)},2,l}\),
corresponds to a hierarchical parameter indexed by \(l=1,\ldots,L\). The
indices for penalization are specified and reordered using internal
mechanisms from \texttt{mgcv::smooth2random} to align with the penalty
structure of the model.

Thus, the full prediction structure is represented as: \[
\eta = \beta_0 + \mathbf{X}_{\texttt{s(x,z)},1}\boldsymbol{\beta}_{\texttt{s(x,z)},1} + \sum_{l} \mathbf{x}_{\texttt{s(x,z)},2,l}^\top\beta_{\texttt{s(x,z)},2,l},
\] where \(\beta_0\) is the intercept,
\(\boldsymbol{\beta}_{\texttt{s(x,z)},1}\) are the parameters for the
unpenalized component, and \(\beta_{\texttt{s(x,z)},2,l}\) are the
parameters for the penalized component.

Using \emph{brms}, these matrices are then transformed into random
effect parameterizations using transformations \(\mathbf{U}\) and
\(\mathbf{D}\). Specifically, \(\mathbf{X}\) is post-multiplied by
\(\mathbf{U}\) and scaled column-wise by \(\mathbf{D}\) where
applicable. The fixed-effect component \(\mathbf{X}_f\) is separated
from random-effect components \(\mathbf{X}_r^{(i)}\), which are indexed
according to the associated penalization structure. The process ensures
that the penalty indices and random effect parameterizations are
correctly ordered.

The fixed effects are combined as
\(\eta_f = \mathbf{X}_f \boldsymbol{\beta}_f\), while the random effects
are represented as
\(\eta_r = \sum_{i} \mathbf{X}_r^{(i)} \boldsymbol{\beta}_r^{(i)}\). The
total predictor is expressed as \(\eta = \beta_0 + \eta_f + \eta_r\),
where \(\beta_0\) is the intercept term. This formulation ensures that
the smooth terms and their random-effect parameterizations are fully
integrated into the regression model, enabling accurate predictions for
new data while maintaining computational efficiency and stability.

\subsubsection*{Supplement C.2: R Code}\label{supplement-c.2-r-code}
\addcontentsline{toc}{subsubsection}{Supplement C.2: R Code}

See website
\url{https://uncertaintree.github.io/research/qoi-check.html}.

\end{document}